\documentclass[twocolumn]{aastex631}

\shorttitle{The uncertainty budget of high-resolution cross-correlation spectroscopy}
\shortauthors{Savel et al.}

\usepackage{xspace}

\begin{document}

\title{Peering into the black box: forward-modeling the uncertainty budget of high-resolution spectroscopy of exoplanet atmospheres}

\newcommand\umd{\affiliation{Astronomy Department, University of Maryland, College Park, 4296 Stadium Dr., College Park, MD 207842 USA}}
\newcommand\cca{\affiliation{Center for Computational Astrophysics, Flatiron Institute, 162 Fifth Avenue, New York, NY 10010, USA}}
\newcommand\umich{\affiliation{Department of Astronomy, University of Michigan, 1085 South University Avenue, Ann Arbor, MI 48109, USA}}
\newcommand\uchicago{\affiliation{Department of Astronomy \& Astrophysics, University of Chicago, Chicago, IL 60637, USA}}
\newcommand\asu{\affiliation{School of Earth and Space Exploration, Arizona State University, PO Box 871404, Tempe, AZ 85281, USA}}

\definecolor{rb4}{HTML}{27408B} 
\newcommand{\kw}[1]{{\color{rb4}[KW: #1 ]}}

\newcommand\addressresponse[1]{\textcolor{purple}{\textbf{#1}}}

\newcommand\addressresponsesecond[1]{\textcolor{teal}{\textbf{#1}}}
\makeatother

\newcommand\name{\texttt{scope}\xspace}

\author[0000-0002-2454-768X]{Arjun B. Savel}
\cca
\umd

\author[0000-0001-9907-7742]{Megan Bedell}
\cca

\author[0000-0002-1337-9051]{Eliza M.-R. Kempton}
\umd

\author[0000-0002-9946-5259]{Peter Smith}
\asu

\author[0000-0003-4733-6532]{Jacob L. Bean}
\uchicago

\author[0000-0002-3852-3590]{Lily L. Zhao}
\cca

\author[0000-0001-8432-7788]{Kaze W.K. Wong}
\cca

\author[0000-0002-9142-6378]{Jorge A. Sanchez}
\asu

\author[0000-0002-2338-476X]{Michael R. Line}
\asu

\begin{abstract}
Ground-based high-resolution cross-correlation spectroscopy (HRCCS; $R \gtrsim 15{,}000$) is a powerful complement to space-based studies of exoplanet atmospheres. By resolving individual spectral lines, HRCCS can precisely measure chemical abundance ratios, directly constrain atmospheric dynamics, and robustly probe multidimensional physics. But the subtleties of HRCCS datasets---e.g., the lack of exoplanetary spectra visible by eye and the statistically complex process of telluric removal---can make interpreting them difficult. In this work, we seek to clarify the uncertainty budget of HRCCS with a forward-modeling approach. We present a HRCCS observation simulator, \name,\footnote{\url{https://github.com/arjunsavel/scope}} that incorporates spectral contributions from the exoplanet, star, tellurics, and instrument. This tool allows us to control the underlying dataset, enabling controlled experimentation with complex HRCCS methods. Simulating a fiducial hot Jupiter dataset (WASP-77Ab emission with IGRINS), we first confirm via multiple tests that the commonly used principal components analysis does not bias the planetary signal when few components are used. Furthermore, we demonstrate that mildly varying tellurics and moderate wavelength solution errors induce only mild decreases in HRCCS detection significance. However, limiting-case, strongly varying tellurics can bias the retrieved velocities and gas abundances. Additionally, in the low-SNR limit, constraints on gas abundances become highly non-Gaussian. Our investigation of the uncertainties and potential biases inherent in HRCCS data analysis enables greater confidence in scientific results from this maturing method.
\end{abstract}

\keywords{Exoplanet atmospheric composition	(2021) --- 
Radiative transfer simulations (1967) --- High resolution spectroscopy (2096) --- Infrared spectroscopy (2285) --- Astronomy data modeling (1859)}

\section{Introduction} \label{sec:intro}
Exoplanet atmospheres offer insight into extreme regimes. When studying the atmospheres accessible with current instruments, we must confront the fundamental properties of gas in high-temperature, low-pressure environments \citep{showman2002atmospheric}. Doing so pushes our understanding of atmospheric dynamics, microphysical processes, and gas-phase chemistry into new territory \citep[e.g.,][]{mbarek2016clouds, powell2018formation, tan2019atmospheric, beltz2021exploring, gao2021aerosols}.

In the past decade, ground-based high-resolution cross-correlation spectroscopy \citep[HRCCS; $R \gtrsim 15,000$;][]{birkby2018exoplanet} has carved out a compelling niche as a probe of exoplanet atmospheres. This technique is unique in its ability to directly constrain exoplanetary winds \citep[e.g.,][]{snellen2010orbital, kempton2012constraining, showman2013doppler}, and by resolving tens of thousands of individual spectral lines, it has enabled precise constraints on chemical composition \citep[e.g.,][]{line2021solar, gandhi2022spatially, maguire2022high, boucher2023co}. With clear evidence of three-dimensional physics and chemistry \citep{flowers2019high, beltz2020significant, ehrenreich2020nightside, wardenier2021decomposing, savel2021no, gandhi2022spatially,beltz2023magnetic,nortmann2024crires} and the potential to unveil dozens of chemical species in the optical and near-infrared \citep[e.g.,][]{carleo2022gaps, kesseli2022atomic, prinoth2022titanium}, HRCCS datasets promise a strong complement to their lower-resolution counterparts, one that will endure even in the era of JWST atmospheric studies \citep{gardner2006james, greene2016characterizing, brogi2019retrieving}.

Complicating the high scientific potential of HRCCS are the subtleties in its analysis. Unlike in low-resolution studies, exoplanet spectra in general are not detected in each observed spectral channel with HRCCS, but are rather buried in the noise and not visible ``by eye.'' Therefore, the presence of the planetary signal in HRCCS must be inferred from cross-correlation functions \citep[e.g.,][]{snellen2010orbital}. This feature of HRCCS is accompanied by ``black box''-like data processing. While the steps to this analysis are fundamentally deterministic and \textit{can} be understood with forward modeling, the complexity of the underlying physical processes (namely tellurics) in turn requires complex statistical treatments. 

Further complicating matters is the axis of researcher choice. Previous studies have sought to shed light on the impact of these processing steps, comprehensively exploring the sensitivity of HRCCS results to analysis choices 
\citep{de2013detection,cabot2019robustness,langeveld2021assessing, gully2022interpretable,meech2022applications,brogi2023roasting, cheverall2023robustness}. Furthermore, studies often ``inject'' additional signal into data to check the responsivity of their dataset to true signal \citep[e.g,.][]{gibson2022relative}. Even so, it is difficult to exactly isolate in which circumstances the data, the analysis techniques, or some combination drives the behavior of the results. Taken together, these aspects have produced an opaque uncertainty budget for HRCCS.

In this work, we seek to provide intuition for the less intuitive aspects of HRCCS data processing and analysis to better understand the HRCCS uncertainty budget. We do so with a forward-modeling approach, simulating HRCCS emission spectroscopy datasets and analyzing them as one would analyze real data. Our focus is threefold: on variable tellurics and their removal, on wavelength solution stability, and on the level of photon noise. All of these uncertainty sources are unexplored in HRCCS from the modeling perspective, and they are hypothesized to explain outstanding questions in HRCCS \citep[e.g., the sub-unity scaling of the planetary spectrum found by][]{van2022carbon}. For this work, our model aims to mimic the HRCCS dataset of WASP-77Ab from \cite{line2021solar} observed with the IGRINS spectrograph \citep{yuk2010igrins, park2014igrins, igrins2017ppl,mace2018igrins}. We focus on this dataset because of its high quality, the availability of many telluric standard stars at this observatory, and the validation of this dataset's analysis by recent JWST results \citep{august2023jwst}. 
Though our work focuses on simulating emission spectra, our results have direct implications for transmission spectroscopy studies as well, and we will comment on them in detail. Section~\ref{sec:methods} describes the methods of this work, including our HRCCS data simulation pipeline---which we name \name---and our reduction of computational burden. Section~\ref{sec:results} contains the results of our forward-modeling experiments. Section~\ref{sec:discussion} discusses our results, including a list of known caveats. Finally, we conclude this work in Section~\ref{sec:conclusions}.

\section{Methods} \label{sec:methods}
\begin{figure*}
    \centering
    \includegraphics[scale=.7]{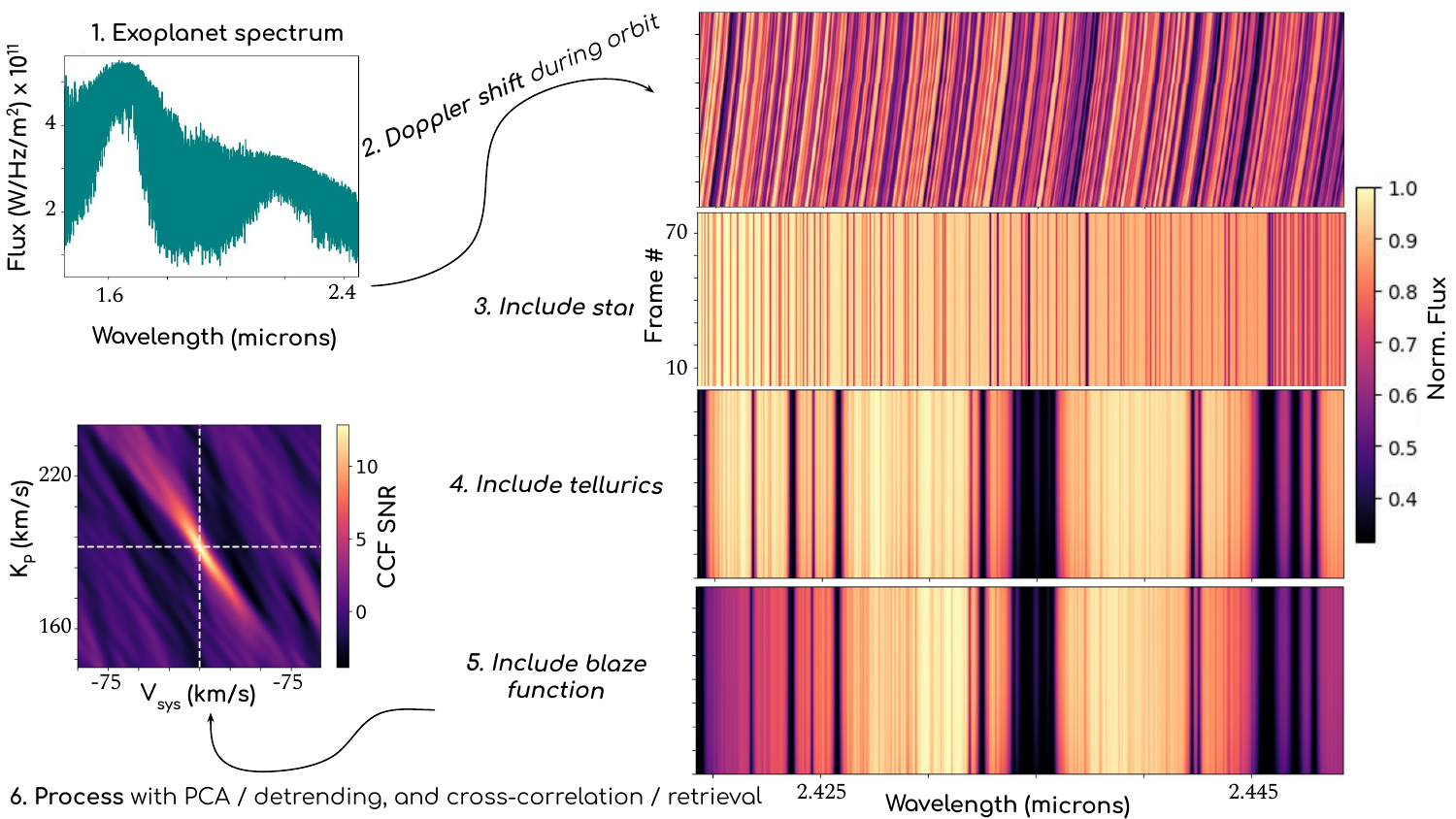}
    \caption{Schematic of \name pipeline. This example uses a planetary emission spectrum calculated with the CHIMERA \citep{line2013systematic} atmospheric model as input. This simulation is meant to reconstruct the IGRINS dataset used to identify the emission of WASP-77Ab by \cite{line2021solar}. The slight discontinuity in the planetary emission tracks is due to a gap in exposures corresponding to these observations. The SNR of this simulation matches the SNR of the  \citealt{line2021solar} observations (SNR=12.8); 9 other noise realizations span SNRs from 11.0 to 19.0. }
    \label{fig:pipeline}
\end{figure*}

To better understand the impacts of HRCCS data-processing steps and choices, we simulate HRCCS observations with the \name pipeline (Fig.~\ref{fig:pipeline}). After detailing the steps involved in simulating these datasets, we then describe the signal extraction procedures used. Afterward, we discuss how we motivate the data-driven tellurics forward model specifically, and finally how we assess spectrograph instability to motivate the instabilities injected in our datasets.
\subsection{Baseline observation simulation pipeline and data processing} \label{sec:baseline_pipeline}
Our forward-modeling approach is similar in spirit to the methods described in \cite{brogi2019retrieving}. To create an HRCCS dataset, \name:
\begin{enumerate}
    \item Calculates time stamps of simulated observations
    \item Creates a data cube
    \item Includes and Doppler shifts planetary spectrum
    \item Includes and Doppler shifts stellar spectrum
    \item Includes telluric absorption
    \item Includes the blaze function
    \item Includes photon noise
    \item Includes wavelength solution instability (ignored in the baseline case)
\end{enumerate}

\subsubsection{Integration time calculator}\label{sec:ETC}
Setting exposure times in HRCCS requires trading off two considerations. Exposing longer decreases the photon noise; however, exposing shorter means the planetary lines have less time to Doppler shift a planetary pixel, or ``smear'' (see \citealt{boldtchristmas2023opt} for an investigation of this tradeoff). 

For this study, we use the exposure times (70 s) and time sequences from \cite{line2021solar}. \name can also calculate the exposure time before a planet's lines cross a pixel. Because IGRINS has 3.3 pixels per resolution element with a resolving power per element of roughly 45,000 \citep[e.g.,][]{mace2018igrins}, the velocity scale of an individual resolution element is $\approx 6.7$~km\,s$^{-1}$, and the velocity scale of a single pixel is roughly $\approx 2.0$~km\,s$^{-1}$. With this value determined, we calculate the maximum integration time before this ``crossing,'' $t_{\rm int}$, as

\begin{equation}\label{eq:integration time}
    t_{\rm int} = \frac{v_{\rm pixel}}{\frac{dv_{\rm LOS, planet}}{dt}},
\end{equation}

for the planetary change in line-of-sight velocity $dv_{\rm planet}/dt$ and the velocity scale of pixel $v_{\rm pixel}$. We use the \texttt{exoplanet} \citep{foreman_mackey_daniel_2021_7191939} package to numerically evaluate the line-of-sight acceleration of the modeled planet at each point in its orbit, and we take the maximum time as the constant exposure time for our set of observations. The maximum integration time we calculate for WASP-77Ab over the simulated phase range (260~s) is much greater than the actual adopted exposure time (70~s). 

It should be noted that overheads should also be considered when setting exposure times. While they do not necessarily impact Doppler smearing, they do factor into the total amount of photons collected during the night and therefore the statistical significance of any recovered signals. The readout time for the WASP-77Ab observations with IGRINS in \cite{line2021solar} is 28~s per frame, which is a significant fraction of the exposure time.

Setting the exposure time sets the time sequence of our simulated observations, which in turn controls the number of observations in our data cube (Section~\ref{sec: data cube}), the Doppler shifting of the planetary (Section~\ref{sec: planet spectrum} and stellar (Section~\ref{sec: stellar spectrum}) spectra, the telluric variation (Section~\ref{sec:pipeline tellurics}), and the signal-to-noise ratio (SNR) that can be achieved for our observations (Section~\ref{sec: noise}).

\subsubsection{Data cube}\label{sec: data cube}
In HRCCS datasets, observed data are customarily packaged in data cubes as, e.g., matrices of shape $N_{\rm exposures} \times N_{\rm orders} \times N_{\rm pixels}$ \citep[e.g.,][though note that $N_{\rm exposures}$ and  $N_{\rm orders}$ are often swapped in the literature; e.g., \citealt{brogi2023roasting}]{line2021solar}. To construct our data cube, we match the number of exposures (79) in the \cite{line2021solar} dataset. We then extract the order and pixel grid (44 orders by 1848 pixels) from the WASP-77Ab dataset. As in \cite{line2021solar}, we drop 10 orders that are severely impacted by tellurics in the IGRINS waveband.

\subsubsection{Planet spectrum}\label{sec: planet spectrum}

In this study, we model planetary emission with the CHIMERA code \citep{line2013systematic, line2014systematic,line2015uniform,line2016no,line2017uniform,line2021solar,line2013systematiciii,brogi2019retrieving}, using the best-fitting atmospheric and velocity parameters from \cite{line2021solar}, calculated at a native resolution of 250,000 prior to convolution. Our temperature--pressure profile is calculated using the \cite{madhusudhan2009temperature} prescription.

To include a planetary spectrum (scaled by the square of the planetary radius) in our simulated observations, we first convolve the spectrum down to the instrumental resolution. We also convolve the spectrum with a rotational kernel to include the effect of planetary rotational broadening \citep[e.g.,][]{gray2021observation}. Finally, we calculate the planetary orbital radial velocity at each time step in our data cube and Doppler shift the planetary spectrum to that velocity. The planetary spectrum must be calculated to regions well outside the instrumental wavelength range, so that Doppler shifting and convolving the spectrum does not introduce spurious edge effects.

For a subset of our simulations, we alter both the planetary and stellar (see \ref{sec: stellar spectrum}) signals to include the observer's movement around the Solar System's barycenter. Doing so introduces another net time-varying velocity component to the measured signal.

\subsubsection{Stellar spectrum}\label{sec: stellar spectrum}
Incorporating a stellar spectrum involves similar convolving and Doppler shifting as the planetary spectrum (Section~\ref{sec: planet spectrum}), the key difference being that the stellar spectrum is Doppler shifted to its radial velocity around its system barycenter, as opposed to the planetary orbital motion. Both the planet and star can be shifted by the system's overall radial velocity, which we set to the 1.6845 ~km\,s$^{-1}$ systemic velocity of WASP-77Ab \citep{maxted2012wasp}.

For the simulations presented in this work, we exclusively use PHOENIX stellar models \citep{husser2013new} scaled by the square of the stellar radius. For our emission benchmark, we use the same PHOENIX stellar model corresponding to the known parameters of WASP-77A \citep[$T_{\rm eff}$ = 5500~K; $\log g$ = 4.33; Fe/H = 0.00;][]{maxted2012wasp}. In the IGRINS waveband (roughly 1.45 to 2.45 microns), these models have a median native resolution of 580,000; like the planetary models, they are then convolved down to instrumental resolution.

\subsubsection{Tellurics}\label{sec:pipeline tellurics}
To include telluric absorption in our simulation, we first determine the airmass at each timestamp of our data cube. For the simulations presented in this work, we use the airmasses from the \cite{line2021solar} observations, which range from 1.5 to 3.1. We then incorporate tellurics by multiplying the data cube by telluric spectrum at each timestamp, interpolating to the wavelength grid of each order. See Section~\ref{sec: data tellurics} for this work's calculation of tellurics.

\subsubsection{Blaze function}
The final spectral component of \name is the instrumental blaze function. Given that this work focuses on reproducing IGRINS observations, we provide a fit specifically to the IGRINS blaze function based on standard star observations.

We access the IGRINS standard star data by scraping the Raw \& Reduced IGRINS Spectral Archive (RRISA).\footnote{\url{https://igrinscontact.github.io/}} We gather observations from 28 nights and search the headers of those observations for the targets that were used as telluric standards. This process yields a total of 93 observations of 4 A0V standard stars.

With our standard star observations acquired, we next determine the blaze function from our observations. In sum, we do so by extracting the IGRINS blaze function on an order-by-order basis with an iterative spline-fitting procedure. To remove spectral features due to absorption lines, we first fit a given order with a low-order spline, then divide the data by the fit. We iteratively remove any flux that is 1.5$\sigma$ below the mean of this fit-divided data until this process converges.\footnote{We find that 1.5$\sigma$ is adequate to drop out the main absorption features while preserving the continuum} We repeat this process until no other points are removed from the dataset. We then fit a tenth-order polynomial to this absorption-cleaned standard star spectrum, which has the flexibility to capture the behavior of the blaze function.

The initial data products from which we fit the blaze function are normalized to unity, so they cannot capture differences in throughput between orders. To capture order-dependent throughput variations, we fit a piecewise second-order polynomial to the throughput variations across orders from the \cite{line2021solar} dataset.

\subsubsection{Noise}\label{sec: noise}
With all spectral components included, we add shot noise per pixel as an optional, final step. We initially determine the signal-to-noise ratio (SNR) that can be achieved with our data cube's time steps using the exposure time calculator for IGRINS \citep{le2015exposure}. This approach assumes a seeing of 0.6$''$ and requires picking a stellar magnitude. For this work, we fix SNR = 250 for one set of simulations (to match the \cite{line2021solar}) and SNR = 60 to investigate a low-SNR limiting case. The lower-SNR case effectively simulates observations of a fainter star with the same integration time.

To capture the throughput of the instrument, we scale our simulated flux cube such that the 90th percentile of the flux cube matches the 90th percentile of the \cite{line2021solar} counts on a per-order basis---that is, each exposure is assumed to have the same throughput. Finally, we simulate a Poisson draw for each pixel to generate our noisy data.

\subsubsection{PCA}
With our HRCCS data cube simulated, the next step to simulating the HRCCS analysis is to perform signal extraction. Principal components analysis (PCA) is a method for abstracting low-dimensional trends from high-dimensional datasets \citep[e.g.,][]{jolliffe2016principal}. In the context of HRCCS, it is used to remove the dominant spectral components---the blaze function, the stellar spectrum, the tellurics---and preserve shot noise and the planetary spectrum \citep[e.g.,][]{line2021solar}. Prior to performing PCA (which we implement with single value decomposition, or SVD), we take standard data-cleaning steps to maintain the statistical validity of the operation, such as sending NaNs to 0, mean-centering the input dataset, and dividing it by its standard deviation. We decompose the data matrix using the \texttt{numpy.linalg} package, setting the first few singular values to zero. The matrix is then reconstructed without those first few singular components, effectively removing the stationary components in the data cube. 

\subsubsection{The cross-correlation function}
With the simulated HRCCS dataset ostensibly cleaned of non-planetary spectral components, \name is now able to compare exoplanet atmospheric models to the simulated dataset. This comparison is fundamentally performed via the cross-correlation function \citep[e.g.,][]{brogi2019retrieving}

\begin{equation}
    \text{CCF}(t, s) = \frac{\Sigma^nF(\lambda, t)m(\lambda-s)}{n\sqrt{v_Fv_m}},
\end{equation}
where $m$ is the template model that is cross-correlated against the data, $v_F$ is the variance of the data, $v_m$ is the variance of the model, $s$ is the amount by which the template model has been shifted, and there are $n$ total data points. The data are here mean-subtracted prior to cross-correlation. Because this is a forward-modeling exercise, there exists a template model that exactly matches our input dataset---which would not be the case for a real planetary atmosphere. Despite the presence of this artificial template match, the forward-modeling exercise is still useful because the data is subjected to the various HRCCS data-processing steps that can sculpt the signal.

The above equation describes the CCF at a single Doppler shift. In general, cross-correlation is performed along an axis---i.e., the resultant CCF is a vector with values corresponding to each Doppler shift of the template model, maximized at the Doppler shift where the model best matches the data. This process allows model comparison to a single observation. HRCCS observations, however, generally involve multiple exposures taken at different orbital phases. The planet therefore has a different line-of-sight velocity in each exposure. HRCCS analyses encapsulate this effect by computing ``$K_{\rm P}$--$V_{\rm sys}$ maps,'' for planetary orbital velocity $K_{\rm P}$ and systemic velocity $V_{\rm sys}$.
We follow the standard literature \citep[e.g.,][]{brogi2012signature, brogi2019retrieving} to compute our $K_{\rm P}$--$V_{\rm sys}$ maps:

\begin{enumerate}
    \item For a given pair of $K_{\rm P}$ and $V_{\rm sys}$, calculate the line-of-sight planetary velocity as a function of orbital phase.
    \item Doppler shift the template model to the expected velocity at each exposure for this orbital solution.
    \item Scale the template with a tunable scale factor (default set to 1) and by the data after removing the contributions from the data’s first $\rm N_PCA$ components
    \item Perform PCA on the scaled template model
    \item Cross correlate each exposure in the data with the corresponding Doppler-shifted template model.
    \item Sum the cross-correlation values for this solution across all exposures to arrive at the value for this $K_{\rm P}$--$V_{\rm sys}$ pair.
\end{enumerate}

Once the map is completed (as a grid spanning $\pm 100$~km\,s$^{-1}$ in $K_{\rm P}$ and $\pm 100$~km\,s$^{-1}$ in $V_{\rm sys}$ around the expected values), we compute detection significance. There exist multiple ways to determine significance from CCF maps, such as iterative sigma clipping \citep{kasper2021confirmation} and the Welsh's T test \citep{brogi2013detection}. In this work, we calculate detection significance simply by subtracting the map by its mean and by dividing by its standard deviation far from the expected peak in the map (with $K_{\rm P}$ spanning 93.06~km\,s$^{-1}$ to 172.06~km\,s$^{-1}$ and $V_{\rm sys}$ spanning $-$100~km\,s$^{-1}$ to $-$11~km\,s$^{-1}$), in order to match the approach of \cite{line2021solar}.

\subsubsection{Retrievals}
The CCF map does not provide quantitative estimates on the properties of a planet's atmosphere. To calculate these, we must turn to Bayesian inference, which in turn requires a mapping to a likelihood function \citep{brogi2019retrieving, gandhi2019hydra, gibson2020detection}. On top of the forward model and statistical sampler required for low-resolution retrievals \citep[e.g.,][]{madhusudhan2019exoplanetary}, high-resolution retrievals require Doppler-shifting of the template model at a given $K_{\rm P}$--$V_{\rm sys}$ pair and model processing (i.e., performing PCA on models in the way that they were performed on data; \citealt{brogi2019retrieving}).

In this work, we perform HRCCS retrievals using the CCF--likelihood mapping derived in \cite{brogi2019retrieving}. To avoid mismatch between the CCF template and the planetary model within the simulated dataset, we also use CHIMERA as our forward model.

For our sampling algorithm, we significantly depart from previous works. Past retrievals have used Markov Chain Monte Carlo \citep[MCMC; specifically, \texttt{emcee};][]{foreman_mackey_daniel_2021_7191939, gibson2020detection, maguire2022high} or nested sampling \citep[specifically, \texttt{\texttt{PyMultiNest}};][]{buchner2014x, brogi2019retrieving, line2021solar} to sample posterior distributions. However, vanilla MCMC can have difficulty in exploring posterior distributions that are strongly degenerate or multimodal \citep[e.g.,][]{madhusudhan2018atmospheric}, and nested sampling algorithms often take long periods of time to converge. In this work, we use the \texttt{flowMC} \citep[][Appendix~\ref{appendix:flowmc}]{wong2022flow} package to accelerate the posterior distribution estimation. This method pairs a local, gradient-based sampler (to effectively characterize local minima) with an adaptive, non-local sampler (to effectively find global minima). Switching to this sampler, in addition to other optimizations \citep[e.g., re-implementing the retrieval in JAX; Appendix~\ref{appendix:jax};][]{jax2018github} decreased runtime from 8 days on 24 CPUs and 1 GPU (with \texttt{\texttt{PyMultiNest}}) to 15 hours on 1 CPU and 1 GPU, or only 5 hours when restricted to local sampling.\footnote{For these performance benchmarks, we used an Nvidia A100-80GB GPU, and we used CPUs on a 4-core Intel Ice Lake node with 1 TB of RAM.}

\subsection{Perturbations to baseline pipeline}
Beyond the baseline modeling of Section~\ref{sec:baseline_pipeline}, we seek in this work to explore higher-order observation-related effects that may appear in HRCCS data. We focus in particular on two such effects: time-varying tellurics and moderate wavelength instability.

\subsubsection{Modeling tellurics}\label{sec: data tellurics}

\begin{figure}
    \centering
\includegraphics[scale=0.45]{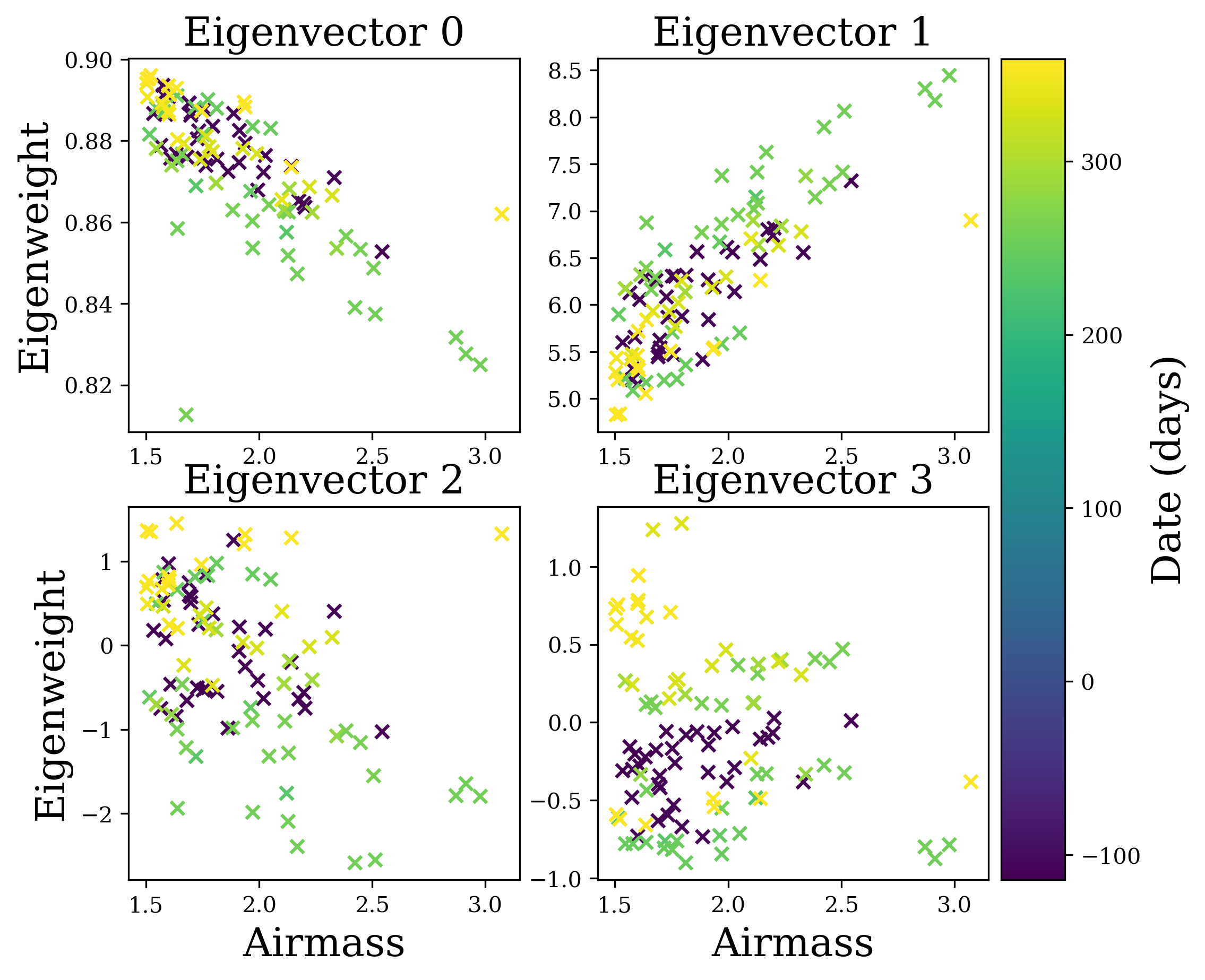}
    \caption{The eigenweights of our PCA decomposition for IGRINS standard star observations for a single order (order 5 in the H band). In this case, the first two components clearly covary with airmass, and multiple components exhibit subtle substructure as as function of date.}
    \label{fig:eigenweights}
\end{figure}

Absorption by the Earth's atmosphere is traditionally modeled out of HRCCS datasets based on synthetic modeling of the Earth's atmosphere \citep[e.g., \texttt{Molecfit};][]{smette2015molecfit,kausch2015molecfit} and/or data-driven approaches \citep[e.g., PCA;][]{giacobbe2021five}. It is unclear how well the synthetic models capture potential time dependence of observed telluric absorption at telescopes, especially with respect to ``microtellurics'' \citep[e.g.,][]{cunha2014tellurics, wang2022characterizing}. It is similarly unclear to what extent the finer details of tellurics are fully removed by existing data-driven methods. These questions can be answered with a forward-modeling approach: by reconstructing telluric absorption using real observations, we can assess its impact on HRCCS data with controlled ``experiments.''

In this work, we construct a data-driven forward model of tellurics by performing PCA on observed telluric spectra. Then, using these PCA components, we build a forward model to simulate the telluric spectrum as a function of airmass and date. The aim of this simplified model is not to accurately predict the telluric spectrum at future times, but only to yield a reasonably true-to-life representation of the way that the telluric spectrum varies in a subset of IGRINS data, so that we can simulate a realistic representation of telluric variability.

To construct a reasonable estimate of telluric absorption as observed by IGRINS, we begin again with our set of A0V standard star observations (as scaled by Vega models\footnote{Upon inspection near the Balmer lines, we find that this Vega model correction imprints limited residuals, resulting in a relatively pure telluric spectrum.}) from the IGRINS Raw and Reduced Spectral Archive. Some nights contain multiple (up to 6) observations; the median difference between exposures is 36 days. We then mean-scale each dataset and perform PCA on the cube of spectra (with dimensions of 93 spectra by 2048 pixels), keeping the first four eigenvectors and eigenweights. These first four components contain a majority (more than 75\%) of our dataset's variance. Capping the components at four risks losing some higher-order telluric variability signals that may still impact the planet detection, but given the quality of telluric standard star data available we didn't want to risk imprinting photon noise by going to higher numbers of components, so this should be regarded as an incomplete test of telluric variability's effects.

From each observation, we extract the spectrum, the airmass, and the timestamp. Humidity and precipitable water vapor (PWV) are known to correlate with singular vectors in HRCCS data analysis; PWV in particular can vary on the order of minutes, potentially driving strong time-variability in HRCCS signals \citep{chiavassa2019planet, smith2024combined}. Rapid changes in PWV are not incorporated in our tellurics model and may cause more complex variability than we account for. In light of this, we also test the impact of the ambient humidity recorded with each observation.

\begin{figure*}
    \centering
    \includegraphics[scale=0.8]{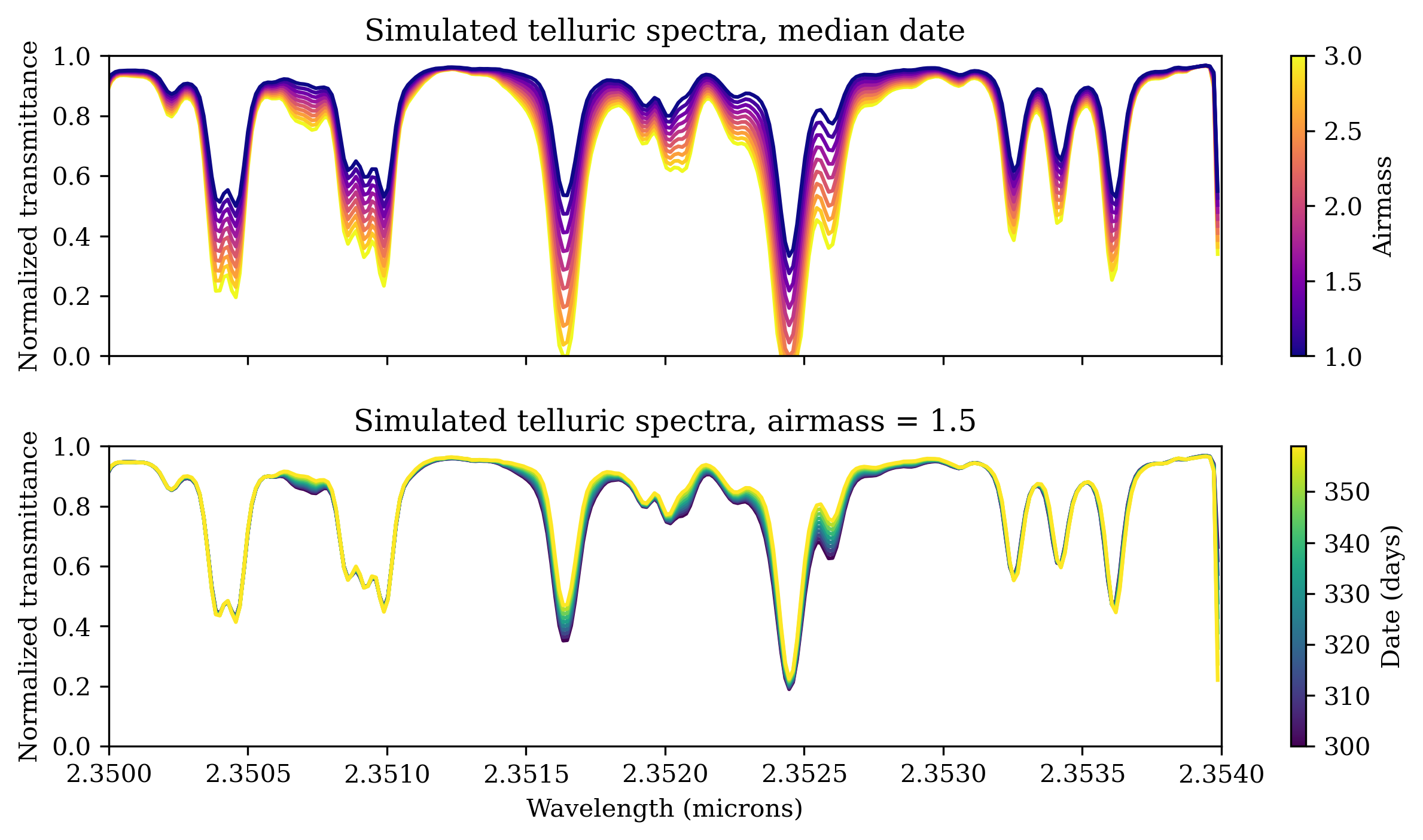}
    \caption{Simulated telluric spectra, zoomed in on a narrow spectral region to display the effect of changing airmass (top) and date (bottom). Airmass is clearly the first-order effect, but some lines are particularly time-variable.}
    \label{fig:sim_telluric}
\end{figure*}

These eigenvectors and corresponding eigenweights describe the primary variation of the telluric absorption at each subsequent exposure. We assume that the primary time-dependence of this absorption is driven by airmass, followed by variations in time. We assess this hypothesis with a linear regression for these data on airmass and time. We leave out 10\% of the initial dataset (12 of 120 spectra) to assess the quality of our regression.

We find that the telluric absorption observed by IGRINS exhibits both a trend in airmass and a trend in time. To select the best representation of eigenweights with airmass and time, we predict the left-out 10\% of spectra and seek to minimize the $\chi^2$ of those predictions. Of the models explored to characterize our eigenweights, the best-performing model scales the eigenweights linearly in airmass and quadratic in time. We do not find that including humidity markedly increases the performance of our model (potentially because of our poor time sampling), so we continue with a model that assumes only airmass and time.

The results of our tellurics forward modeling experiment are shown in Figs.~\ref{fig:eigenweights}--\ref{fig:sim_telluric}. The first two eigenweights in particular exhibit linear trends in airmass. The corresponding shifts in the contributions of the first two eigenvectors with airmass produce deepening absorption lines at higher airmass.

While Fig.~\ref{fig:eigenweights} shows light clustering in eigenweights as a function of time, it is not immediately evident from this figure how the full spectral model changes with time. The spectrum-level changes with time are more evident in Fig.~\ref{fig:sim_telluric}, with two classes of spectral regions emerging: static lines and lines that vary on the scale of months. We verify that the two strongest lines with time variability in the plotted wavelength region are in regions of strong water opacity \citep{polyansky2018exomol}.

As an extension of our tellurics model, we additionally include a forward model that lengthens the modeled timescale (one night) by a factor of 100. This approach effectively increases the variability of the tellurics and approximates a multi-night observing campaign.

\begin{figure}
    \centering
\includegraphics[scale=0.55]{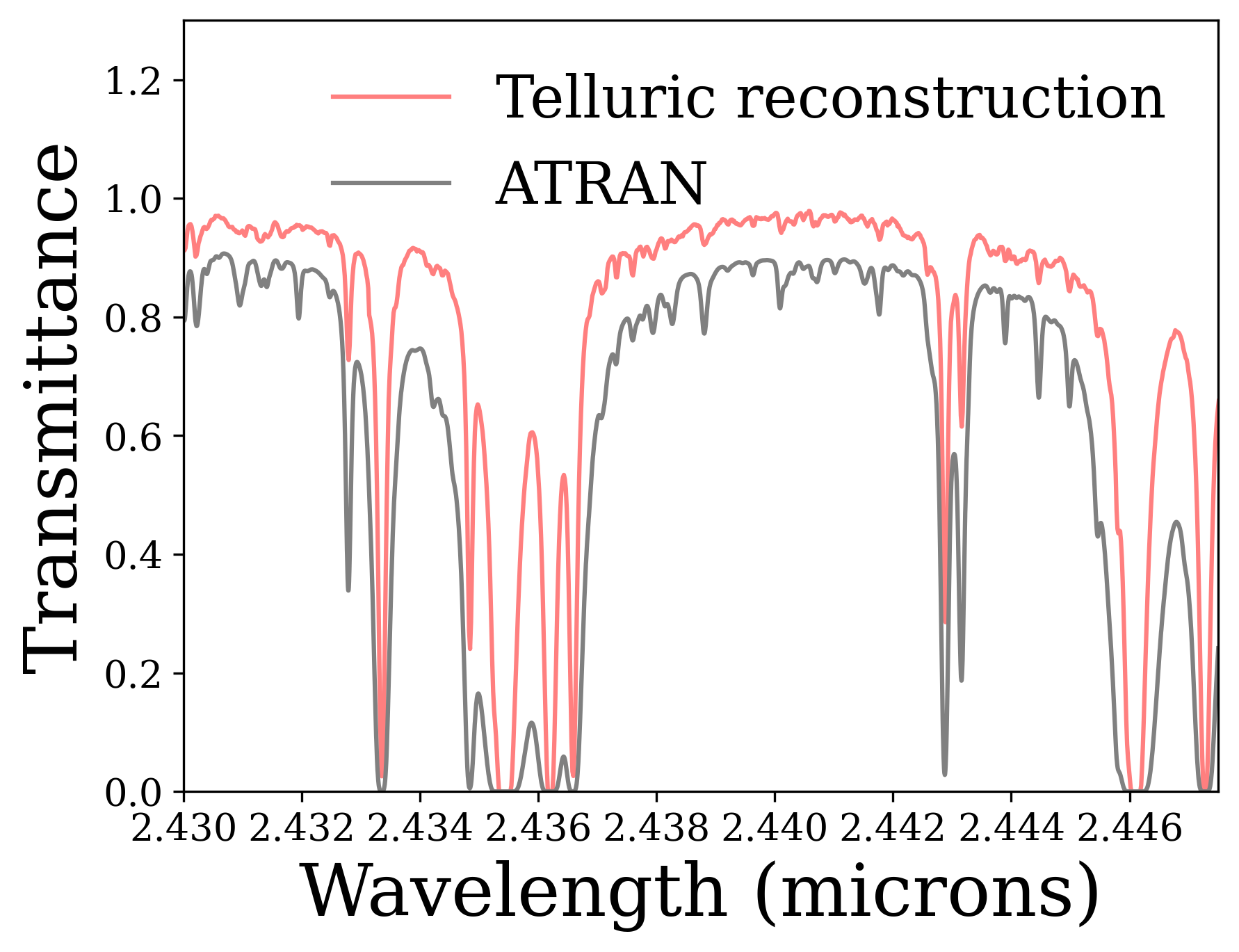}
    \caption{A comparison of our data-driven telluric reconstruction against ATRAN \citep{lord1992new} radiative transfer simulations over one IGRINS order.}
    \label{fig:atran_compare}
\end{figure}

Our data-driven approach broadly matches the line positions and depths of simulations with ATRAN (\citealt{lord1992new}; Fig.~\ref{fig:atran_compare}), a radiative transfer telluric modeling code that takes as input the altitude, latitude, water vapor overburden, and zenith angle of observations. Differences in the absolute continuum level are likely due to our approach of rescaling the telluric standard stars prior to performing PCA. Because the continuum level of the flux is lost by HRCCS data-processing \citep[e.g.,][]{brogi2019retrieving}, these differences in the continuum between telluric approaches are not expected to impact inference.

\subsubsection{Telluric removal: airmass-detrending and Molecfit}
Beyond our baseline PCA procedure for telluric-removal, we explore alternative telluric-removal methods to understand the sensitivity of our results to such approaches.

Detrending datasets by airmass has a long history in HRCCS \citep[e.g.,][]{snellen2010orbital}. We implement it in this work by fitting a second-order polynomial to our dataset to each pixel as a function of time, motivated by the expected change in line depth as the telescope points through longer columns of air. More robust approaches in the literature generally follow this step with higher-order regression on the residuals of the fit and filtering to remove additional substructure \citep[e.g.,][]{brogi2012signature}. While our limited approach requires less tuning and is adequate to understand changes in significance relative to changes in simulated data, the approach is likely not reliable enough to be applied on real datasets because we do not perform additional filtering steps.

To further assess the impact of PCA on detection significance, we performed tests where we apply matrix standardization prior to the PCA, multi-linear regression afterward, and dividing out the PCA fit instead of subtracting \citep[e.g.,][]{brogi2023roasting}. Doing so emphasizes relative errors (therefore overweighting, e.g., tellurics).

Finally, we test correction of telluric absorption with \texttt{Molecfit}. In contrast with the more empirical approach of airmass-detrending, \texttt{Molecfit} models the absorption of the Earth's atmosphere by fitting radiative transfer-derived spectra to (subsets of) data. Based on the absorption noted in \cite{smette2015molecfit}, we fit our \texttt{Molecfit} runs with respect to $\rm H_2O$, $\rm CH_4$, and $\rm CO_2$. When available, we pin input values (e.g., with relative humidity) to values recorded by the \cite{line2021solar} IGRINS observations. Otherwise, we resort to \texttt{Molecfit} defaults (e.g., with ambient temperature).

\subsubsection{Stability of wavelength solution}\label{sec: wav_solution_fit}
Unstabilized spectrographs, such as IGRINS, are susceptible to changes in the wavelength solution over the course of a night. For instance, \cite{brogi2023roasting} report wavelength solution variability of $\pm 0.46$~km\,s$^{-1}$ over the course of their IGRINS observations. As described in Section~\ref{sec:baseline_pipeline}, \name can incorporate inaccuracies in the wavelength solution to determine their effect on inference from HRCCS data.  
We implement this effect by incorporating an additional Doppler shift to our data cube, unrelated to astrophysical sources.

To motivate the functional form and degree of wavelength instability in our model, we assess the wavelength instability of the \cite{line2021solar} dataset. This instability found in the dataset will then be included in our HRCCS forward models. We seek in this dataset a simple shift and drift (i.e., a changing shift in time) of the wavelength solution in time as recorded by IGRINS. To assess this variation, we try two approaches: fitting for the position of a single line (confirmed non-stellar spectral feature that occurs across all observations), and for the position of all telluric lines simultaneously by cross-correlating against an ATRAN telluric model \citep{lord1992new}. We chose this pair of approaches to explore limiting cases (i.e., considering shifts of one line vs. all lines at once).

Our line-fitting experiment indicates that, over the course of a single night, IGRINS data exhibit significant shifts in wavelength solution. Our cross-correlation drifts are on the order of 0.3~km\,s$^{-1}$, whereas the single-line drifts are on the order of 0.1~km\,s$^{-1}$. These shifts are on the order of a tenth of a pixel in resolution, and therefore may play a role in subsequent cross-correlation analysis. The similarity between these two drift estimates over time indicates that we have identified a real effect in these data. Notably, this level of precision is on the order of the wavelength solution precision reported by \cite{brogi2012signature} on CRIRES \citep{ulrich2008crires}, using a wavelength solution approach similar to \cite{line2021solar}. When injecting wavelength instability into the pipeline model, we adopt a drift of 0.3~km\,s$^{-1}$.

\section{Results} \label{sec:results}
\subsection{$K_{\rm P}$--$V_{\rm sys}$ maps}
We begin by presenting the $K_{\rm P}$--$V_{\rm sys}$ forward-modeling maps, as these data products are the simplest scientifically meaningful units that \name produces. These maps indicate the significance of the planetary detection and can additionally reveal spurious signals.

\subsubsection{Varying the number of PCA components}\label{sec:remove_PCA}

\begin{figure}
    \centering
    \includegraphics[scale=0.55]{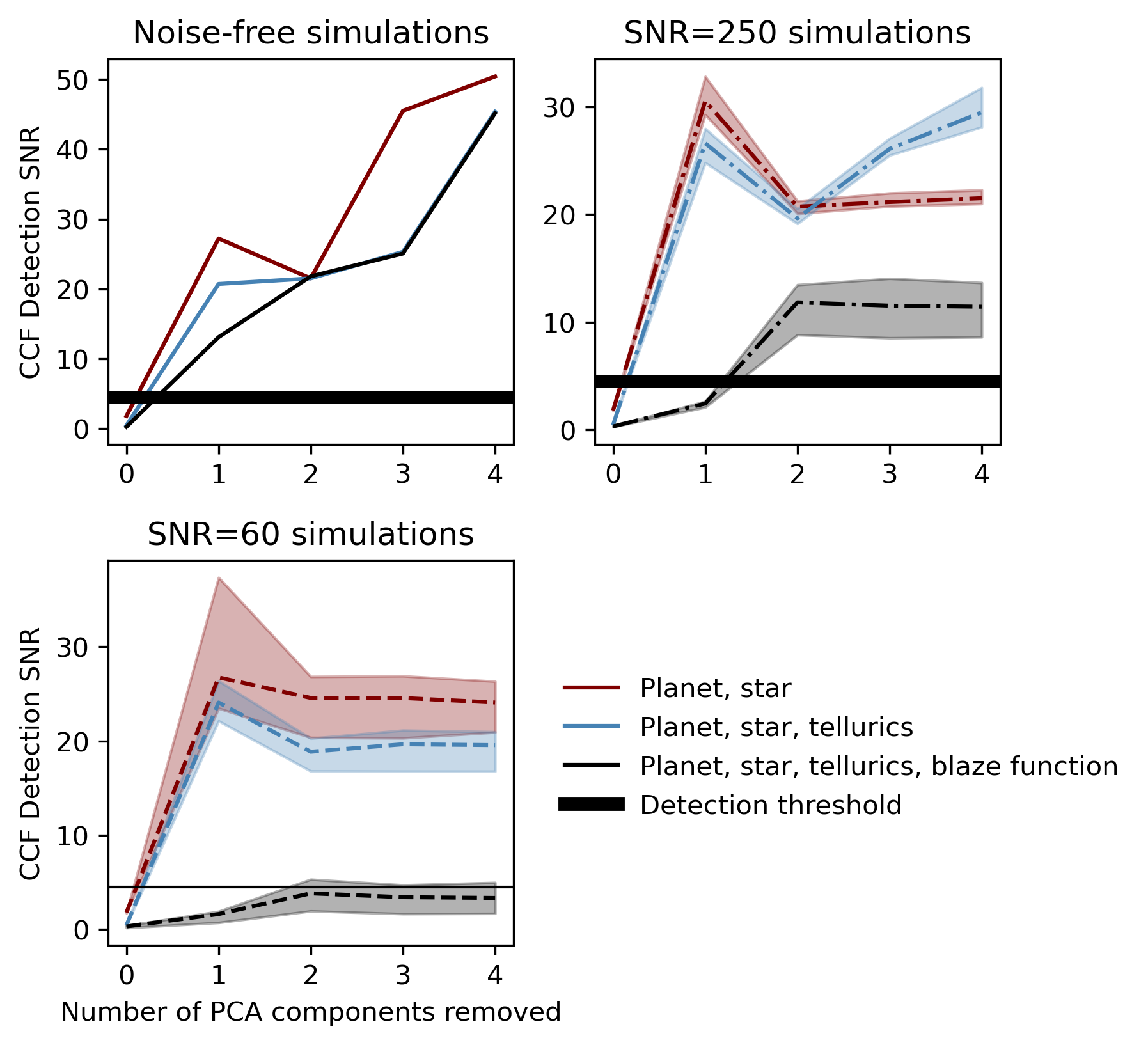}
    \caption{The effect of removing different numbers of PCA components from HRCCS data on the CCF detection significance in our baseline results. This is calculated in noiseless simulations (top left), SNR = 250 simulations (top right), and SNR = 60 simulations (bottom left) with 10 different noise realizations. Including more spectral components (e.g., the stellar spectrum) severely decreases the planet's detection significance, but a small number of PCA components can recover the planet detection about the detection threshold. These significance values are compared to a commonly accepted HRCCS detection threshold \citep[SNR=4.5; e.g.,][]{cabot2019robustness}.}
\label{fig:pca_removal_res}
\end{figure}
In this section, we report how the planetary spectrum detection significance changes in the presence of different levels of noise to test its sensitivity to PCA treatment.

Fig.~\ref{fig:pca_removal_res} shows the effect of increasing the number of PCA components on the CCF detection significance from 0 to 4. Of first note is that cross-correlating the planetary spectrum against a pure planetary data cube---an unphysical test case with no star, tellurics, or blaze function---yields a very high-significance detection even when the photon noise level is relatively high (i.e., in the SNR=60 simulation). 

Once more spectral components are included, the detection significance of the planetary atmosphere is severely diminished, as those components have features that dominate over planetary ones. In all cases, at least one PCA component is required to bring the detection significance above the detection threshold \citep[commonly held to be about SNR=4.5;][]{cabot2019robustness}.

\begin{figure*}
    \centering
    \includegraphics[scale=0.7]{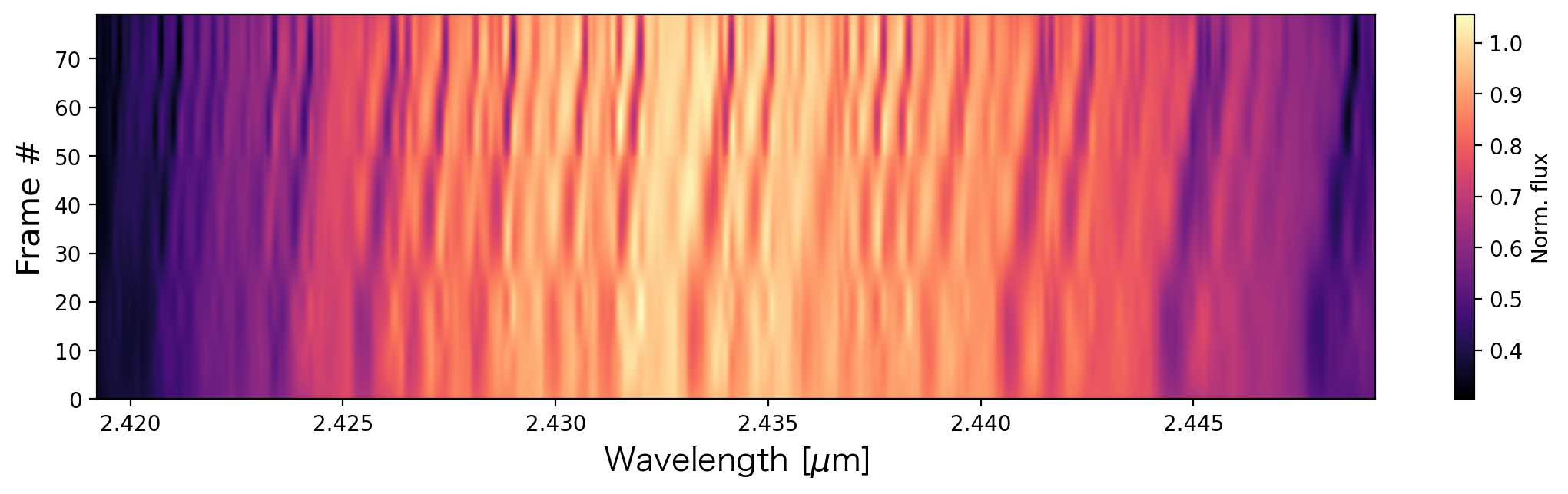}
    \caption{The spectral components learned by four PCA components from a simulated HRCCS dataset that only includes the planetary signal and instrumental blaze function. The PCA algorithm is able to approximate the diagonal trace of the planetary atmosphere with a superposition of a few strong stationary components. When other spectral components are present (including photon noise), PCA is not able to approximate the planetary signal, as it is most sensitive to the time-invariant signals with greater variance across wavelengths.}
    \label{fig:beat_pattern}
\end{figure*}

When modeling just the planet and star combination without noise, the detection significance begins to diminish when more than 1 PCA component is used. This reduction of the planetary detection significance is possible even though PCA cannot capture the moving planetary signal as a single component. As we show in Fig.~\ref{fig:beat_pattern}, with enough freedom and without many other spectral components, multiple components can together approximate the planetary signal as a sum of stationary components. This pattern, however, is not present in a realistically noisy case (SNR=250).

While removing a low number of PCA components does not severely impact planetary detection, we do see reduced detection significance for a very high numbers of removed components. In particular, we can produce spurious non-detections when many components are removed; that is, as the PCA algorithm is provided more flexibility, it can create artificial correlated structures (peaks and troughs) in the $K_{\rm P}$--$V_{\rm sys}$ maps. As a limiting example, when we remove the maximum possible number of PCA components (79), the planet signal entirely disappears.  Despite these dangers, the experiments presented here indicate that using a reasonably low number of PCA components on realistically noisy data does not meaningfully lessen the detection significance. 

\begin{figure}
    \centering
    \includegraphics[scale=0.39]{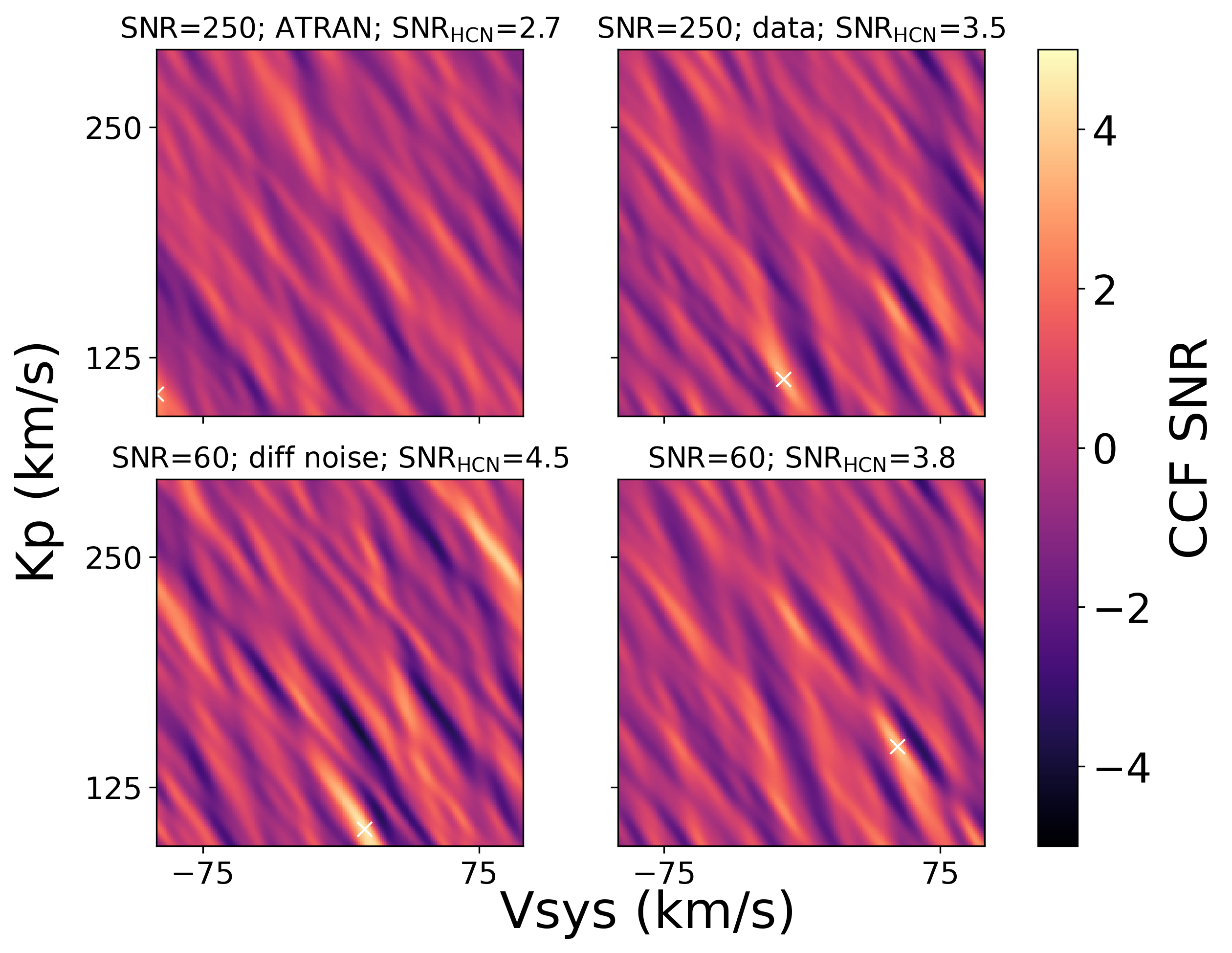}
    \caption{A set of spurious HCN signals in our baseline simulations. The maxima of these $K_{\rm P}$--$V_{\rm sys}$ are denoted with a white ``x.'' While HCN is not included in this set of models, its signal can be found} in cross-correlation when using ATRAN tellurics (top left), data-driven tellurics (top right), with low photon noise (top row), with high photon noise (bottom row), and in different noise realizations (bottom left).
    \label{fig:false_hcn}
\end{figure}

\subsubsection{Instrumental and telluric variability}\label{sec: kp vsys time dep tellurics}

We next tested two effects that could potentially interfere with atmospheric characterization with HRCCS: variability in the instrumental wavelength solution and variability in the tellurics. 

By forward-modeling the wavelength instability that we recover from IGRINS observations, we find that such instability introduces essentially no decrease in detection significance in the low-noise case and a 0.5 SNR decrease in detection significance in the high-noise case. Similarly, our time-variability exercise indicates that the tellurics as we model them do not vary enough within a given night to impact HRCCS observations. The smooth changing of our line depths is simply not rapid enough to induce strong changes in $K_{\rm P}$--$V_{\rm sys}$ maps. As a limiting-case investigation, our longer-timescale telluric variability yields a decrease in SNR of 0.8.

\subsubsection{Barycentric motion}\label{sec: res barycentric motion}
In the high-SNR regime (photon SNR=250), we find that including barycentric motion reduces the CCF SNR by about 1 on average. In the noiseless regime, there is negligible difference between the two; in the low-SNR regime (photon SNR=60), there is also negligible difference between the two.

\subsubsection{Spurious signals}\label{sec:result spurious}

Our results are able to produce lower-significance signals related to species that are not included in the forward model. We focus in particular on HCN, motivated by the idea that the gas's intrinsic line strength and periodicity in wavelength might make it more susceptible to spurious signals than other species \citep{zhang2020platon}. To test, this, we generate a dataset that includes all best-fitting values of species from \cite{line2021solar} aside from HCN. Despite its exclusion in the forward model, \name measures a spurious HCN signal at a significance of SNR=$3.5$ in the low-noise case, albeit at the wrong $K_{\rm P}$ value (Fig.~\ref{fig:false_hcn}). Note that this peak is matched by a trough of roughly the same magnitude in this simulation.

We also note spurious HCN signals when using different sets of tellurics (trading the ATRAN radiative transfer tellurics for our data-driven tellurics forward model), different noise levels (in both the low and high photon noise case), and in different noise realizations. The structure of the $K_{\rm P}$--$V_{\rm sys}$ varies when the tellurics model and the noise instance change, and the strength of the  spurious signal changes with all model changes. The  spurious signal is robust to telluric treatment and noise instance. Additionally, the $K_{\rm P}$--$V_{\rm sys}$ structure in a spurious signal scenario is altered by both the noise instance and tellurics simulation.

\subsection{Likelihood sampling}\label{sec: kp vsys likelihood}

We have shown (Section~\ref{sec:remove_PCA}) that the planetary detection significance is severely weakened when other spectral components are introduced to a simulated HRCCS dataset. 

As a next step, we move beyond the $K_{\rm P}$--$V_{\rm sys}$ maps by optimizing the likelihood function over those two velocity terms and a linear scale factor, $a$. The scale factor \citep[e.g.,][]{pino2020neutral, van2022carbon, brogi2023roasting} stretches the planetary spectrum, which may accommodate associated stretching from data processing steps such as PCA. This likelihood sampling approach is analogous to a full atmospheric retrieval, except that all planetary atmospheric parameters are held constant---only the velocity terms and scale factor are fit here. This task is relatively computationally intensive; we perform our fitting with the \texttt{emcee} \citep{foreman_mackey_daniel_2021_7191939} package, parallelizing across 25 CPUs with MPI to avoid consuming GPU resources.

We find that including four PCA components does not impact the retrieved scale factor (Fig.~\ref{fig:scale_factor_fit}). Our result is consistent at both noise levels considered, with the width of the posteriors simply increasing by a factor roughly equal to the noise increase (roughly 3).

\subsection{Retrievals}
As the final step in our suite of experiments, we perform full atmospheric retrievals on our simulated datasets to assess whether any biases are introduced.

\subsubsection{Baseline comparison to \cite{line2021solar}}
Given the complexity of the algorithms used, we firstly benchmark our simulations against the \cite{line2021solar} observational results to validate \name.

Broadly, we are able to reproduce the basic results of \cite{brogi2019retrieving}: the retrieval framework adequately recovers the input parameters. Furthermore, our modeled parameter constraints are about as precise as the \cite{line2021solar} constraints. For instance, our constraints on water ($-4.08^{+0.16}_{-0.13}$ dex) and $V_{\rm sys}$ ($0.09^{+0.81}_{-0.86}$ ~km\,s$^{-1}$) approximately match the corresponding water ($-3.93^{+0.10}_{-0.09}$ dex) and $V_{\rm sys}$ ($-7.71^{+5.27}_{-5.41}$ ~km\,s$^{-1}$) constraints from \cite{line2021solar} in both value and uncertainty (Fig.~\ref{fig:gas_constraints} with an 8~km\,s$^{-1}$ offset applied in our simulations).

\subsubsection{Comparison to likelihood sampling approach}
Our baseline retrieval shows a departure from the results in Section~\ref{sec: kp vsys likelihood}: while the scale factor is still consistent with 0, there is an asymmetry in the scale factor posterior (Fig.~\ref{fig:scale_factor_retrieval}). This asymmetry coincides with the degeneracy between this parameter and two of the temperature--pressure (TP) profile parameters that are themselves asymmetric.

\begin{figure}
    \centering
    \includegraphics[scale=0.45]{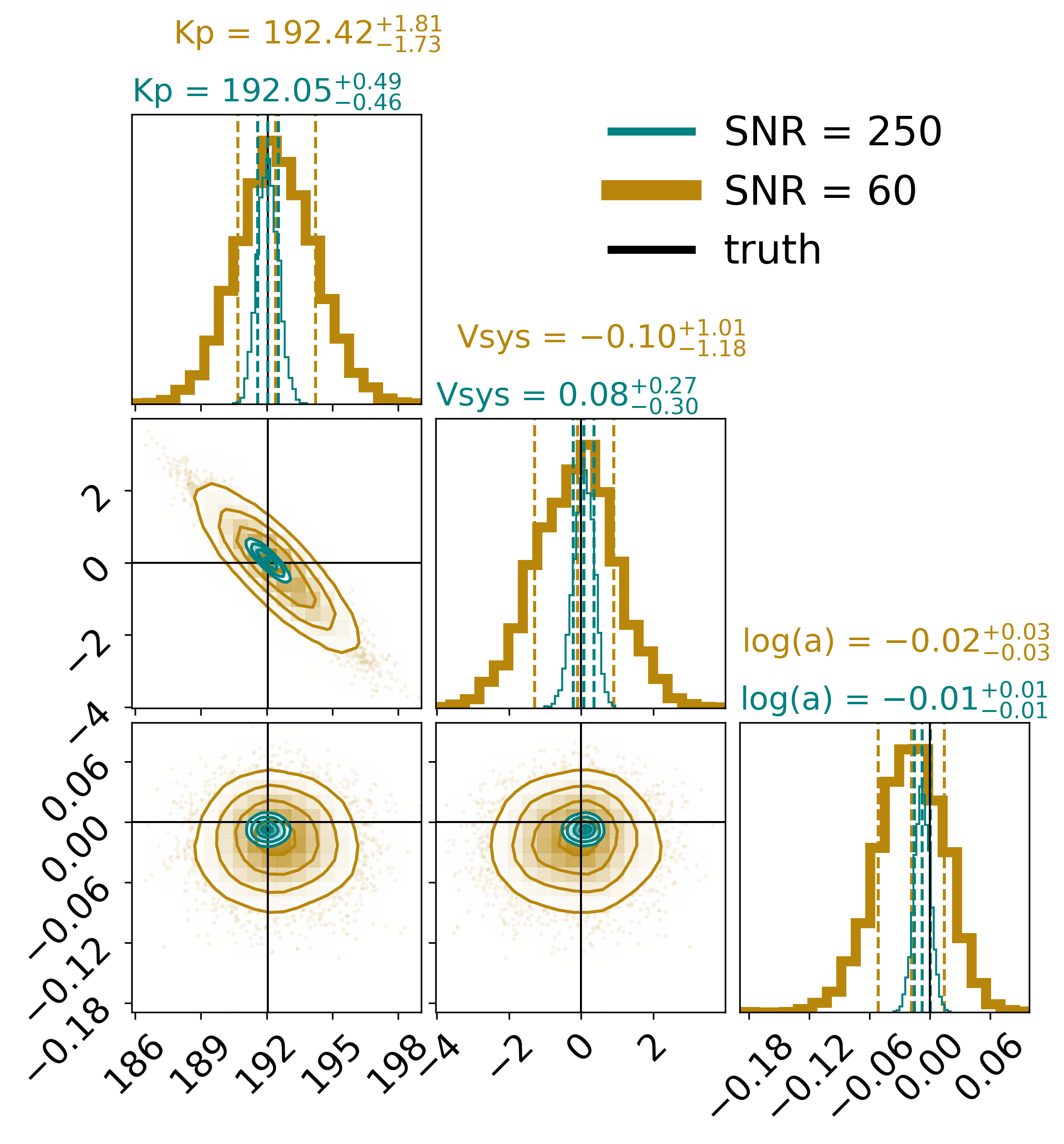}
    \caption{Results of an MCMC fit to the velocity parameters (in~km\,s$^{-1}$) and scale factor for simulated HRCCS data at two photon noise levels. This model used four PCA components to remove the stellar, telluric, and instrumental features in the simulated data. Using PCA in this experiment did not bias the scale factor toward smaller values, indicating that using PCA alone does not scale the planetary signal in HRCCS data.}
    \label{fig:scale_factor_fit}
\end{figure}

\subsubsection{Sensitivity to variable tellurics}
Our previous experiment in Section~\ref{sec: kp vsys time dep tellurics} showed that including time-dependent tellurics decreases the significance of a planetary detection. How does such an effect impact retrieved quantities?

\begin{figure}
    \centering
    \includegraphics[scale=0.35]{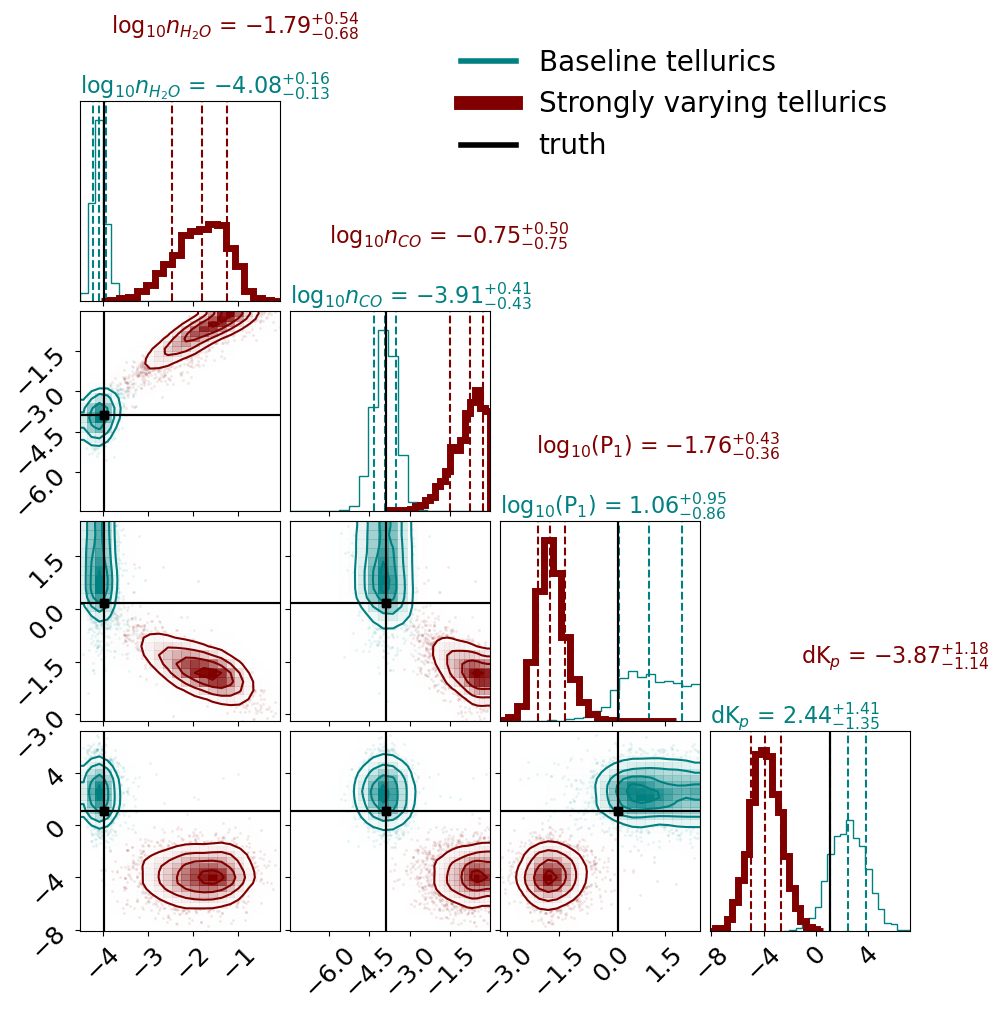}
    \caption{Gas and velocity constraints from our \texttt{\texttt{PyMultiNest}} high-resolution atmospheric retrieval. Including strongly time-dependent tellurics (maroon) yields distinct posteriors from our time-dependent telluric (teal) model.}
    \label{fig:scale_factor_retrieval}
\end{figure}

When using our longer-timescale telluric variability, the abundances of our retrieved emission spectra are correctly recovered, but other quantities are strongly biased. The $K_P$ is biased to $-3.87^{+1.18}_{-1.14}$~km\,s$^{-1}$ from $2.44^{+1.41}_{-1.35}$ ~km\,s$^{-1}$ (Fig.~\ref{fig:scale_factor_retrieval}). The first log pressure point is similarly biased to $-1.76^{+0.43}_{-0.36}$ from $1.06^{+0.95}_{-0.86}$. Finally, gas parameters that were previously bound---water and carbon dioxide---become non-Gaussian and/or unbound.

\begin{figure*}
    \centering
\includegraphics[scale=0.4]{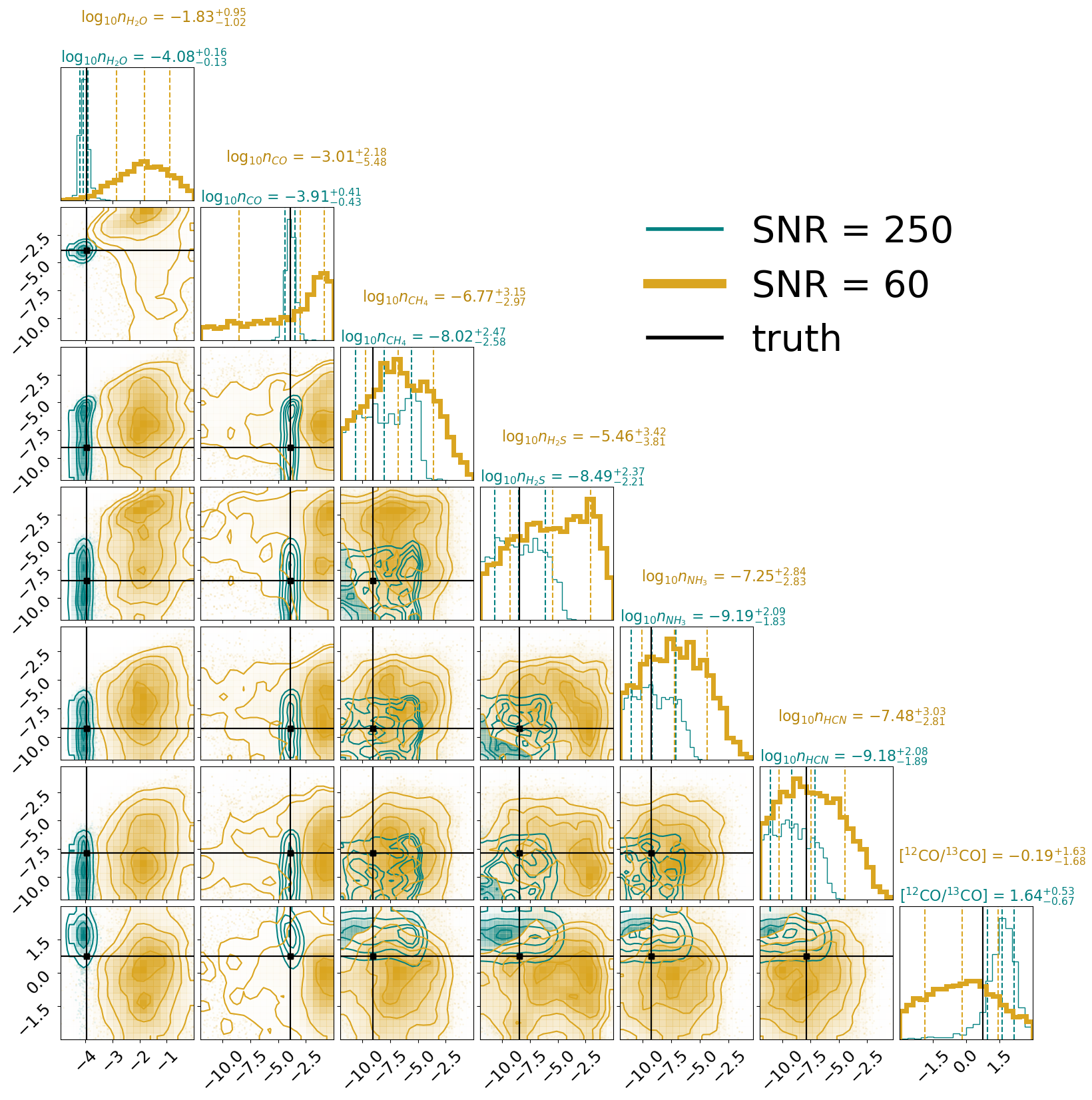}
    \caption{Gas constraints from our baseline retrieval on simulated data of WASP-77Ab. We are able to recover gas constraints about as well as \cite{line2021solar}, with bounded constraints on $\rm H_2O$, CO, and the CO isotope ratio in the lower-noise case (SNR=250). The higher-noise case (SNR=60) substantially changes the likelihood surface for the gas constraints shown, with less of an impact on velocity and temperature--pressure parameters.}
    \label{fig:gas_constraints}
\end{figure*}

\subsubsection{Sensitivity to noise levels}
When we increase the noise in our modeled HRCCS datasets, it appears 
that the gas constraints (Fig.~\ref{fig:gas_constraints}) are more affected by increased photon noise than the velocity constraints are (Fig.~\ref{fig:scale_factor_fit}). While the posteriors on the velocity parameters simply become broader with increased noise, the posteriors of the gas abundances become significantly warped---and in some cases, they become unbounded when they were previously bounded (i.e., with $\rm H_2O$ and CO) and become confidently incorrect (i.e., with HCN).

\section{Discussion} \label{sec:discussion}
\subsection{Confirmation of HRCCS principles}
As evidenced by our $K_{\rm P}$--$V_{\rm sys}$ maps, likelihood sampling, and retrieval results, our simulation reassuringly confirms the basic premise of previous HRCCS proofs-of-concepts. The planetary signal can be recovered in cross-correlation despite its intrinsic spectral weakness; some stellar and telluric removal is required to do so; and the velocity of the planet is imprinted on the data.

The robustness of our results persists even in the presence of mild wavelength instability. This behavior may be expected; the instability measured occurs on the sub-pixel level, so the net effect is to smear out the planetary signal at the sub-pixel level. 

\subsection{Noise and time-varying tellurics can strongly affect HRCCS inference}

While our baseline, more optimistic models reaffirm HRCCS analysis techniques, we identify two potential complicating factors: noise and time-varying tellurics.

\subsubsection{On noise}

Extrapolating from the $K_{\rm P}$--$V_{\rm sys}$ maps and likelihood sampling experiments, it would appear that increasing the amount of photon noise simply broadens the uncertainty on all inferred parameters. However, a full retrieval reveals that the high photon noise case gives poor constraints on gas abundances. 
Additionally, when photon noise is increased, the upper bound on HCN becomes less informative. In light of the retrieval results, the standard for what constitutes an ``acceptable amount of noise'' therefore appears to be higher for retrievals than for the other HRCCS data products.

One reason for the differing gas and velocity posteriors under noise could be that the velocity parameters are sensitive to the \textit{position} of spectral lines, which photon noise does not change, whereas the gas abundances are sensitive to relative line ratios and shapes \citep[e.g.,]{brogi2019retrieving,gandhi2023retrieval}, which are perturbed by increased photon noise. Additionally, \textit{all} the planetary lines contain the planet's bulk velocity information (aside from differences due to atmospheric winds and rotation), whereas inferences from gas abundances of additional species are restricted to subsets of the planetary lines. Therefore, while low-SNR targets may be detected in cross-correlation and may have accurately measured velocities, measuring gas abundances in this regime proves more challenging.

\subsubsection{On time-varying tellurics}\label{sec: discuss time-dep tellurics}

Our $K_{\rm P}$--$V_{\rm sys}$ maps demonstrated negligible differences for our standard time-varying tellurics, and these differences were still modest when using the longer-timescale telluric variability. Similar experiments with airmass-detrending and \texttt{Molecfit} yield stronger deviations (from SNR=10 to SNR=2 for the former). 

The exact reason for these differences is unclear. We modeled our time-dependent tellurics using PCA from the start. Therefore, the PCA-based telluric-removal method may be able to identify and remove this substructure more effectively because of the shared PCA method. Alternatively, the other two approaches as we implement them may be fundamentally less well suited for removing telluric absorption.\textit{This experiment indicates that realistic tellurics likely do not impact IGRINS observations in previously unaccounted-for ways---but that higher telluric variability can decrease detection significance.}

In terms of ramifications, the majority of HRCCS emission studies analyze data from a single night, and they therefore would not be strongly affected by mildly varying tellurics. Additionally, multi-epoch HRCCS observations that aim to keep the planetary spectral lines constant and remove all other sources as variable \citep[e.g.,][]{lockwood2014near,buzard2020simulating, finnerty2021contrast} would in principle be unaffected, as they can treat tellurics as a time-variable process to be removed. However, combinations of multiple nights of data with strongly shifting planetary spectral features may be susceptible to additional error from even mildly varying tellurics over a long enough baseline. This issue could potentially be addressed by performing telluric removal for a given night at a time, then summing the CCF values after the fact \citep[e.g.,][]{kasper2021confirmation}.
 
As described in Section~\ref{sec:results}, we observe a strong bias in temperature--pressure parameters under longer-timescale telluric variability. The reason for this result may be that the retrieval algorithm seeks to compensate for improperly removed telluric signals by mutating the planetary spectrum to account for those spectral components' residual spectral features. This framing also accounts for the $K_{\rm P}$ bias---the presence of spectral features from the star and tellurics at a different velocity from the planet significantly biases the planetary $V_{\rm sys}$. \textit{This experiment demonstrates that the velocity parameter and gas abundances can be biased by tellurics that are not fully removed;} the same intuition should apply to transmission spectroscopy.

While we are able to assess the effect of time-dependent tellurics on the scale of days, a caveat is that our observations do not have a short enough cadence to capture minute- or hour-scale telluric variability (Section~\ref{sec:caveats}).

\subsection{Effects of PCA on the planetary signal}

When all realistic spectral components are included, increasing the number of PCA components does not \textit{decrease} the planetary significance for up to four PCA components. 
While the noise-free, pure-planet case can be learned by PCA (Fig.~\ref{fig:beat_pattern}), simply adding a competing noise term disrupts the coherence of the planetary signal enough so that the planetary signal does not leak into the higher-variance PCA terms.

In our telluric-dominated near-infrared regime, when we perform matrix standardization and multi-linear regression, our signal sizes decrease, with a different number of PCA components now preferred. We anticipate that this approach would not strongly affect retrievals given the reprocessing of the model.

Furthermore, when holding atmospheric parameters constant, the scale factor is fit to be Gaussian around 1 with four PCA components. While the atmospheric signal is certainly stretched by the PCA operation, our result implies it is not stretched in a way that cannot be corrected by model reprocessing in this regime. \textit{This experiment indicates that PCA alone does not reduce CCF significance when using a low number of principal components.}

However, the limiting-case PCA application (with 79 components removed) shows that removing too many PCA components at some point 
becomes detrimental to the planetary signal. This result indicates that as demonstrated by, e.g., \cite{cheverall2023robustness}, there exists an \textit{optimal} number of PCA components to remove for a given dataset. While removing some PCA components is beneficial, removing too many clearly prohibits detecting the planet. This point is expected from the perspective of the definition of PCA---together, all the PCA components explain all the variance in a dataset, so at some point much of the planetary signal will be encompassed by some of the PCA components. In our set of simulations, the optimal number of PCA components appears to be 4. This number, however, is specific only to our simulations as they are modeled, and real data with more complexity (e.g., additional sources of systematic uncertainty) may require a larger number of components to be removed. Again, this finding motivates the search for metrics that characterize the optimal number of components to remove \citep[e.g.,][]{cheverall2023robustness}.

\subsection{On spurious results}
Our modeling results are able to produce spurious signals for HCN at a different $K_{\rm P}$ than input (Section~\ref{sec:result spurious}). A significant signal, even when none is injected, appears across different specific tellurics models and noise realizations, but its location can vary. Because of this, it seems that  spurious signals are a result of a combination of genuine line confusion (i.e., the cross-correlation template picking up on telluric signal) and specific noise instances.

Unfortunately, preventing this spurious signal would not be as simple as limiting the retrieval to the expected $K_{\rm P}$. Given the known strong correlation between these two parameters, strongly limiting one without biasing the other may be difficult. Furthermore, the expected $K_{\rm P}$ for a given observation may simply be incompletely known.

Instead of restricting $K_{\rm P}$ to a single value, one potential way to address off-$K_{\rm P}$ spurious signals would be to enforce a wide uniform prior on $K_{\rm P}$ centered on the expected value, thereby penalizing strongly differing $K_{\rm P}$. The width of this prior could be determined based on the maximum deviation expected from rotation alone \citep{wardenier2023modelling}, with an additional allowance for measurement uncertainty in $K_{\rm P}$. This approach can be readily incorporated into retrieval frameworks, and it may be possible to self-consistently include them in $K_{\rm P}$--$V_{\rm sys}$ map searches for individual gases. 

It may be that our spurious signals are driven by poorly corrected tellurics. If so, masking out these pixels \citep[e.g.,][]{boucher2023co,kanumalla2024igrins} may decrease the impact of this noise source on detection significance. As a test, we mask out the noisiest pixels in our cross-correlation step. We determine which pixels to mask via the inverse survival function of the normal distribution,  \cite{brogi2023roasting}. Doing so allows adaptive thresholding based on the noise characteristics of the dataset. We verify that even when we mask out these selected pixels, we can generate spurious signals of HCN up to a significance of 4.4 in our baseline set of simulations (that is, without the strongly varying tellurics).

It may be that HCN is more likely than other species to generate spurious signals, because of its few, strong, wavelength-periodic lines \citep{zhang2020platon}. Robustly testing this hypothesis would be instructive for interpreting HRCCS inferences and is a promising avenue for future work---especially in the context of the potential tension between HRCCS and JWST inferences of HCN in the atmosphere of the planet HD 209458 b \citep{hawker2018evidence, giacobbe2021five, xue2024jwst}.

\subsection{Caveats}\label{sec:caveats}
While our work incorporates many of the expected first-order astrophysical and instrumental sources in HRCCS datasets, we do not include an exhaustive set. Our limitations can be grouped into five camps:

\begin{itemize}
    \item \textit{limited data reduction and processing steps}. There are steps that are often required to turn a raw data product into science-ready HRCCS data that we do not implement here. For instance, we do not perform bad pixel correction, throughput correction, masking, or high-pass filtering \citep[e.g.,][]{van2022carbon,brogi2023roasting}. We choose not to implement these processes here because they are not assumed to be primary drivers of error in HRCCS data analysis and they are more instrument-specific.
    \item \textit{telluric observation cadence}. We are only able to assess the longer-term variability in tellurics, as the observations of telluric standard stars are not on a short enough cadence to capture, e.g., minute-scale variability. This gap motivates dedicated observing programs for measuring short-timescale telluric variability.
    \item \textit{astrophysical simplicity}. We have furthermore not considered effects such as the multidimensionality of the planetary atmosphere, which strongly sculpts HRCCS data \citep{kempton2012constraining,showman2013doppler, zhang2017constraining, flowers2019high, beltz2020significant, wardenier2021decomposing, savel2021no}, and higher-order effects including condensation, non-local thermodynamic equilibrium, and stellar variability. The broader question of model uncertainty, though certainly intertwined with HRCCS datasets, is out of scope for this work.
    \item \textit{generalizability.} This work models one planet observed by one instrument. The dataset that we compare our simulations to, furthermore, provided a relatively high SNR detection and clear inference. Hence, the simulations here may be more optimistic than they should be in a general case for other planets observed with other instruments. Additionally, while the trends in significance as a function of, e.g., number of PCA components may generalize to other datasets, the exact values will of course not.
    \item \textit{model-on-model simulations.} Our approach inherently assumes an underlying match between the simulated dataset and the forward spectral model. This approach yields unrealistically high detection significances because in real observations, the planetary spectrum is likely not a perfect fit for any model. Furthermore, our simulations include no correlated noise, and other simulation work indicates that red noise and more complex throughput calculations can disrupt the expected scaling of SNR with $\sqrt{N_{\rm lines}}$ \citep[e.g.,][]{, birkby2018exoplanet, lopez2019optimizing, currie2023there}.  

    \item \textit{separation of telluric lines and planetary lines.}  Based on the planetary orbit, the emission spectra that we simulate are well separated from the telluric spectral features. This separation would not be the case for transmission spectra, in which the radial planetary velocity intersects with the stellar velocity, and the planet’s velocity may change comparatively less over the course of observations. Therefore, simulations of HRCCS transmission spectra may yield different results from those shown here, motivating future work.
\end{itemize}

\section{Conclusions}\label{sec:conclusions}
The complex data representation of high-resolution cross-correlation spectroscopy can make it difficult to understand the uncertainty budget of an inference. In this work, we aim to clarify some of more uncertain aspects of HRCCS data processing and analysis. With \name, a newly developed forward model for HRCCS datasets, we find that:
\begin{itemize}
    \item We are able to reduce the computational burden of retrieval algorithms by an order of magnitude through implementing numerical speed-ups with JAX and more efficient statistical sampling with \texttt{flowMC}.
    \item Under our baseline approach, standard HRCCS data processing algorithms work as expected. In our modeling, using a reasonable number (4) of PCA components does not destructively impact the planet signal. 
    \item These results are also robust against mild tellurics as we infer that IGRINS observes them within a single night. This is additionally the case for wavelength instabilities at the level evident in IGRINS data after standard corrections have been performed.
\end{itemize}

Pushing our forward model beyond the baseline cases reveals some HRCCS failure modes. 

\begin{itemize}
    \item While the planetary velocity parameters are robust against increased photon noise, gas constraints deteriorate strongly into non-Gaussian posteriors. 
    \item When strongly time-varying tellurics are improperly removed, gas abundances are biased. The velocity parameters are impacted in this case as well, yielding discrepant Doppler shifts. This bias may arise for multi-epoch observations if each epoch is not cleaned separately.
    \item We search for and are able to produce spurious HCN signals at an incorrect planetary $K_P$ at up to SNR=4.4.
\end{itemize}

The full likelihood surface of these types of observations remains fundamentally underexplored. Further investigations of the dependence of HRCCS constraints on observational setup, analysis techniques, and more complex physics will be required to completely flesh out the strengths and limitations of this powerful technique. Doing so will allow more involved planning and interpreting of HRCCS datasets. By open-sourcing \name, we hope to provide the community with a tool that will enable these investigations.

\begin{acknowledgments}

We thank the anonymous referee for their very thoughtful and helpful comments on this work.

The Flatiron Institute is a division of the Simons Foundation.

A.B.S. and E.M.-R.K. acknowledge funding from the
Heising-Simons Foundation.

The authors thank the Flatiron Institute's Scientific Computing Core for repeated assistance with software package management.

We also thank Emily Rauscher for helpful comments that improved the quality of this manuscript.

This research has made use of NASA’s Astrophysics Data System Bibliographic Services.

\vspace{55pt}

This work used The Immersion Grating Infrared Spectrometer (IGRINS), which was developed under a collaboration between the University of Texas at Austin and the Korea Astronomy and Space Science Institute (KASI) with the financial support of the US National Science Foundation under grants AST-1229522, AST-1702267 and AST-1908892, McDonald Observatory of the University of Texas at Austin, the Korean GMT Project of KASI, the Mt. Cuba Astronomical Foundation and Gemini Observatory.

The authors acknowledge the University of Maryland supercomputing resources (\url{http://hpcc.umd.edu}) made available for conducting the research reported in this paper.

\end{acknowledgments}

\facilities{IGRINS (Gemini South)}

\software{astropy \citep{astropy:2018}, CHIMERA \citep{line2013systematic}, corner \citep{corner}, \texttt{exoplanet} \citep{foreman_mackey_daniel_2021_7191939}, JAX \citep{jax2018github}, flowMC \citep{wong2022flow}, ipython \citep{perez2007ipython}, scipy \citep{virtanen2020scipy}, tqdm \citep{da2019tqdm}, numpy \citep{2020NumPy-Array}, pandas \citep{mckinney2010data}, matplotlib \citep{hunter2007matplotlib}, numba \citep{lam2015numba}}

\appendix
\section{Optimization}\label{sec: time optimization}
The simulations required for HRCCS retrievals are computationally intensive. At its core, the issue is that comparing simulations to datasets from modern echelle spectrographs with large spectral grasps requires solving the radiative transfer equation at hundreds of thousands of wavelength points. Previous studies combated this the runtime issue by sampling a representative wavelength range within the instrumental band \citep[e.g.,][]{kempton2012constraining,wardenier2021decomposing}, with simple parallellization over multiple CPUs \citep[e.g.,][]{savel2021no}, or by parallelizing on GPUs \citep{line2021solar,lee20223d}. The parallelization advances in particular have taken advantage of the fact that the radiative transfer problem can cast as embarassingly parallel over wavelength (or wavelength packet, for Monte Carlo approaches).

To make the HRCCS retrievals in this work computationally feasible enough for repeated use and testing, we undertook an effort to decrease the retrieval code's computational runtime. Across all optimizations, we decreased the runtime from 8 days, 24 CPUs, and 1 GPU to 15 hours, 1 CPU, and 1 GPU.

\subsection{Optimization with JAX}\label{appendix:jax}
JAX \citep{jax2018github} is a code transformation framework in Python. It provides a number of features that enable faster statistical sampling, including automatic differentiation \citep[e.g.,][]{margossian2019review}, just-in-time (JIT) compilation, and GPU acceleration. Because of the code team's main focus being machine learning on accelerators such as GPUs, JAX does not trivially allow common control flows such as if statements. This feature adds nontrivial development cost to port codes to the framework. Previous studies have used JAX for retrievals of exoplanet atmospheres \citep{kawahara2022autodifferentiable}; to our knowledge, this work marks the first time that JAX has been used in HRCCS retrievals.
Using JAX alone successfully decreased the retrieval runtime, but it also allowed gradient-based sampling approaches that reduced the number of samples required for retrieval convergence (see Appendix~\ref{appendix:flowmc}).

Other speed-ups were realized in the process of porting the code to JAX. For instance, we adjusted the code to use linear interpolation of Doppler-shifted spectra via \texttt{numpy} instead of a \texttt{scipy} spline fit; this step was was necessary for the JAX migration because JAX does not support implicit tracing of external libraries.

Previously, the likelihood evaluation of a single atmospheric model was originally CPU parallelized across exposures (i.e., across planetary orbital phases); however, the JAX optimization is fast enough that it can be done without resorting to explicit paralellization with, e.g., \texttt{joblib}. The underlying radiative transfer calculations on the GPU also incurred a 3x speedup, but this is marginal compared to the total evaluation time of a single likelihood call. 

One caveat to the JAX speed-up of this code is the compilation time, which can run in excess of an hour. This cost is negligible, however, in the context of runtimes that generally exceed days. Recent JAX development indicates that compiled code can be cached and reused, so this compilation cost will likely not be a major concern in the future.

Another unique aspect of JAX is that it may copy the memory of global objects. To prevent this behavior while allowing functions to access memory that is only read in once, we partial the opacities into the function definitions themselves.

\subsection{Using \texttt{flowMC}}\label{appendix:flowmc}
\texttt{flowMC} \citep{Wong:2022xvh} is a JAX-based statistical sampling package that couples a local (often gradient-based) sampler with a non-local sampler. Because this work (to our knowledge) marks the first usage of exoplanet atmospheric retrievals with \texttt{flowMC} and a normalizing flow non-local sampler, we detail in this section a set of tests that we conduct to verify the utility of these techniques for our science case.

\texttt{flowMC} essentially adds a non-local proposition to a Markov chain Monte Carlo \citep{gelman2004bayesian} proposal step in conjunction with the  standard local step. While local-stepping is known to converge, reaching a converged state can heavily consume compute resources in the presence of nontrivially shaped posterior distributions \citep{gelman1992inference,tierney1994markov, neal2003slice}. This non-local proposition is designed to learn the complex structure in the posterior geometry; this learning is achieved on the fly with a normalizing flow. Normalizing flows are a machine learning technique---a type of generative model that can be sampled \citep[e.g.,][]{kobyzev2020normalizing,papamakarios2021normalizing}. This model can also be \textit{used} to compute the likelihood of its samples. They operate by performing a change of variables to map from the current sampler location to another posterior mode---the target distribution. This process improves the acceptance of MCMC proposals as the sampling continues \citep[e.g.,][]{gabrie2022adaptive}.  When combined with a gradient-based sampler, this sampling scheme is well-suited for high-dimensional, multimodel sampling problems \citep[e.g,.][]{betancourt2017conceptual,mangoubi2018does,wong2023fast}. 

The critical intuition is that, even a normalizing flow that is not trained perfectly will allow walkers to ``teleport'' from one mode to another, with the ``teleportation'' probability weighted by the target mode's probability mass. Hence, normalizing flows are able to quickly construct the relative mass of modes in a posterior, at the cost of alternating between sampling the posterior and training the normalizing flow on new samples.

\begin{figure}
    \centering
    \includegraphics[scale=0.2]{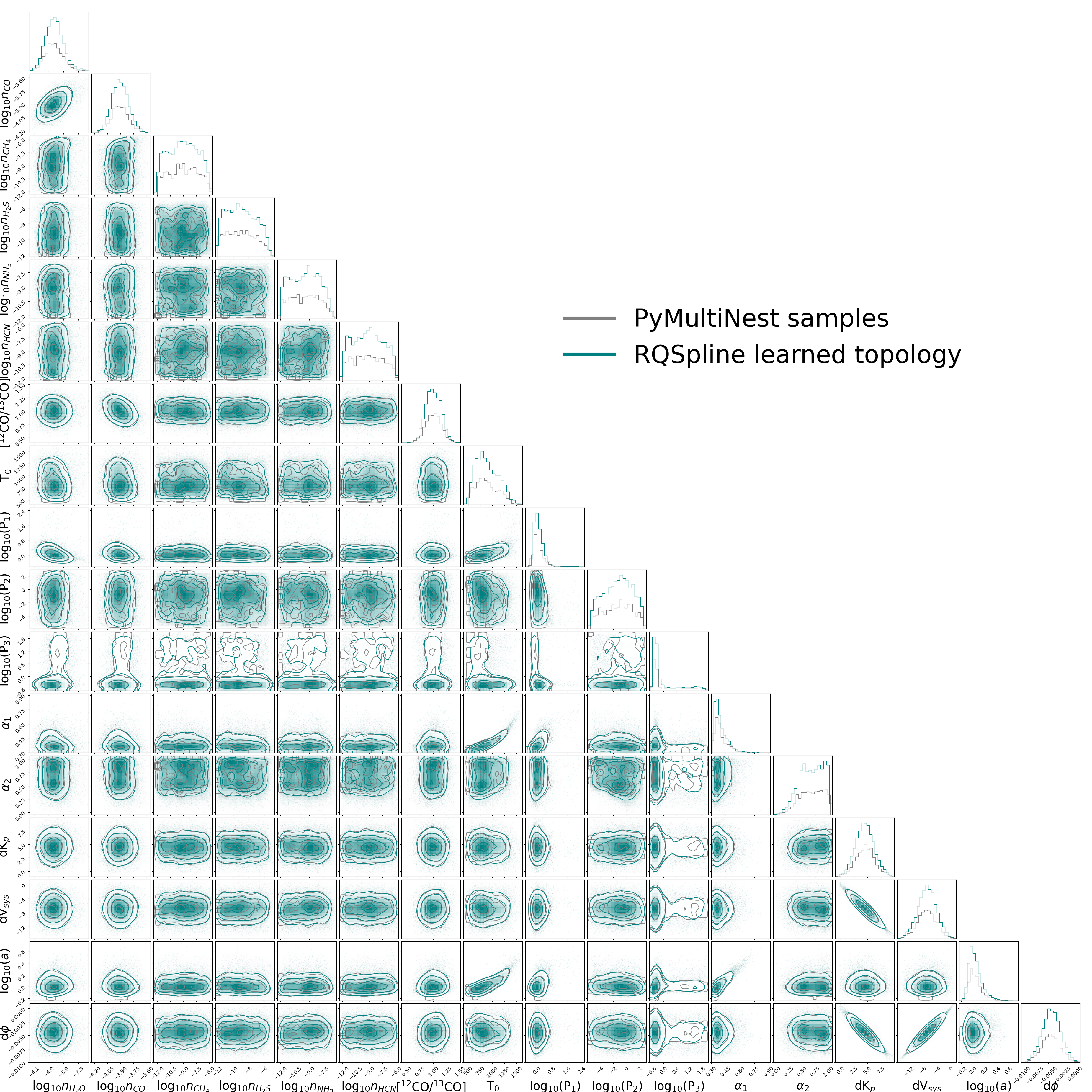}
    \caption{The learned normalizing flow topology as compared with \texttt{PyMultiNest} equally weighted samples from a baseline retrieval. The normalizing flow is able to learn the topology of the posterior distribution. Note that this exercise does not constitute actual parameter inference---rather, we are demonstrating the flexibility of the global sampling scheme to approximate the final global posterior distribution.}
    \label{fig:nf_lean}
\end{figure}

First, we test that our normalizing flow can approximate the topology of a converged \texttt{PyMultiNest} run. For this study, we make use of the RQSpline normalizing flow architecture. With this check successfully verified (Fig.~\ref{fig:nf_lean}, Moving on to retrievals: We find that running \texttt{flowMC} produces posterior distributions that match the \texttt{PyMultiNest} posteriors. Given that our parameters can span different orders of magnitude (e.g., the water abundance vs. temperature at base of the atmosphere), we tune the step size (via the ``mass matrix'') of our local MALA sampler \citep{grenander1994representations} to be inversely proportional to the gradient of the likelihood at the initialization point. While this approach is not optimal, it does well enough to ensure that that correlations between parameters of different magnitudes are well explored. Furthermore, we rescale our top of atmosphere temperature so that it is on the same scale as our abundances (on the order of 10), as opposed to its true range (on the order of 1000).

\bibliography{main}{}

\begin{thebibliography}{}
\expandafter\ifx\csname natexlab\endcsname\relax\def\natexlab#1{#1}\fi
\providecommand{\url}[1]{\href{#1}{#1}}
\providecommand{\dodoi}[1]{doi:~\href{http://doi.org/#1}{\nolinkurl{#1}}}
\providecommand{\doeprint}[1]{\href{http://ascl.net/#1}{\nolinkurl{http://ascl.net/#1}}}
\providecommand{\doarXiv}[1]{\href{https://arxiv.org/abs/#1}{\nolinkurl{https://arxiv.org/abs/#1}}}

\bibitem[{{August} {et~al.}(2023){August}, {Bean}, {Zhang}, {Lunine}, {Xue},
  {Line}, \& {Smith}}]{august2023jwst}
{August}, P.~C., {Bean}, J.~L., {Zhang}, M., {et~al.} 2023, arXiv e-prints,
  arXiv:2305.07753, \dodoi{10.48550/arXiv.2305.07753}

\bibitem[{Beltz {et~al.}(2020)Beltz, Rauscher, Brogi, \&
  Kempton}]{beltz2020significant}
Beltz, H., Rauscher, E., Brogi, M., \& Kempton, E. M.-R. 2020, The Astronomical
  Journal, 161, 1

\bibitem[{Beltz {et~al.}(2023)Beltz, Rauscher, Kempton, Malsky, \&
  Savel}]{beltz2023magnetic}
Beltz, H., Rauscher, E., Kempton, E., Malsky, I., \& Savel, A. 2023, arXiv
  preprint arXiv:2302.13969

\bibitem[{Beltz {et~al.}(2021)Beltz, Rauscher, Roman, \&
  Guilliat}]{beltz2021exploring}
Beltz, H., Rauscher, E., Roman, M.~T., \& Guilliat, A. 2021, The Astronomical
  Journal, 163, 35

\bibitem[{Betancourt(2017)}]{betancourt2017conceptual}
Betancourt, M. 2017, arXiv preprint arXiv:1701.02434

\bibitem[{Birkby(2018)}]{birkby2018exoplanet}
Birkby, J.~L. 2018, arXiv preprint arXiv:1806.04617

\bibitem[{{Boldt-Christmas} {et~al.}(2023){Boldt-Christmas}, {Lesjak},
  {Wehrhahn}, {Piskunov}, {Rains}, {Nortmann}, \&
  {Kochukhov}}]{boldtchristmas2023opt}
{Boldt-Christmas}, L., {Lesjak}, F., {Wehrhahn}, A., {et~al.} 2023, arXiv
  e-prints, arXiv:2312.08320, \dodoi{10.48550/arXiv.2312.08320}

\bibitem[{Boucher {et~al.}(2023)Boucher, Lafreni{\'e}re, Pelletier,
  Darveau-Bernier, Radica, Allart, Artigau, Cook, Debras, Doyon,
  {et~al.}}]{boucher2023co}
Boucher, A., Lafreni{\'e}re, D., Pelletier, S., {et~al.} 2023, Monthly Notices
  of the Royal Astronomical Society, stad1247

\bibitem[{Bradbury {et~al.}(2018)Bradbury, Frostig, Hawkins, Johnson, Leary,
  Maclaurin, Necula, Paszke, Vander{P}las, Wanderman-{M}ilne, \&
  Zhang}]{jax2018github}
Bradbury, J., Frostig, R., Hawkins, P., {et~al.} 2018, {JAX}: composable
  transformations of {P}ython+{N}um{P}y programs, 0.3.13.
\newblock \url{http://github.com/google/jax}

\bibitem[{Brogi \& Line(2019)}]{brogi2019retrieving}
Brogi, M., \& Line, M.~R. 2019, The Astronomical Journal, 157, 114

\bibitem[{Brogi {et~al.}(2013)Brogi, Snellen, De~Kok, Albrecht, Birkby, \&
  De~Mooij}]{brogi2013detection}
Brogi, M., Snellen, I., De~Kok, R., {et~al.} 2013, The Astrophysical Journal,
  767, 27

\bibitem[{Brogi {et~al.}(2012)Brogi, Snellen, De~Kok, Albrecht, Birkby, \&
  De~Mooij}]{brogi2012signature}
Brogi, M., Snellen, I.~A., De~Kok, R.~J., {et~al.} 2012, Nature, 486, 502

\bibitem[{Brogi {et~al.}(2023)Brogi, Emeka-Okafor, Line, Gandhi, Pino, Kempton,
  Rauscher, Parmentier, Bean, Mace, {et~al.}}]{brogi2023roasting}
Brogi, M., Emeka-Okafor, V., Line, M.~R., {et~al.} 2023, The Astronomical
  Journal, 165, 91

\bibitem[{Buchner {et~al.}(2014)Buchner, Georgakakis, Nandra, Hsu, Rangel,
  Brightman, Merloni, Salvato, Donley, \& Kocevski}]{buchner2014x}
Buchner, J., Georgakakis, A., Nandra, K., {et~al.} 2014, Astronomy \&
  Astrophysics, 564, A125

\bibitem[{Buzard {et~al.}(2020)Buzard, Finnerty, Piskorz, Pelletier, Benneke,
  Bender, Lockwood, Wallack, Wilkins, \& Blake}]{buzard2020simulating}
Buzard, C., Finnerty, L., Piskorz, D., {et~al.} 2020, The Astronomical Journal,
  160, 1

\bibitem[{Cabot {et~al.}(2019)Cabot, Madhusudhan, Hawker, \&
  Gandhi}]{cabot2019robustness}
Cabot, S.~H., Madhusudhan, N., Hawker, G.~A., \& Gandhi, S. 2019, Monthly
  Notices of the Royal Astronomical Society, 482, 4422

\bibitem[{Carleo {et~al.}(2022)Carleo, Giacobbe, Guilluy, Cubillos, Bonomo,
  Sozzetti, Brogi, Gandhi, Fossati, Turrini, {et~al.}}]{carleo2022gaps}
Carleo, I., Giacobbe, P., Guilluy, G., {et~al.} 2022, The Astronomical Journal,
  164, 101

\bibitem[{Cheverall {et~al.}(2023)Cheverall, Madhusudhan, \&
  Holmberg}]{cheverall2023robustness}
Cheverall, C.~J., Madhusudhan, N., \& Holmberg, M. 2023, arXiv preprint
  arXiv:2303.01496

\bibitem[{Chiavassa \& Brogi(2019)}]{chiavassa2019planet}
Chiavassa, A., \& Brogi, M. 2019, Astronomy \& Astrophysics, 631, A100

\bibitem[{{Cunha} {et~al.}(2014){Cunha}, {Santos}, {Figueira}, {Santerne},
  {Bertaux}, \& {Lovis}}]{cunha2014tellurics}
{Cunha}, D., {Santos}, N.~C., {Figueira}, P., {et~al.} 2014, \aap, 568, A35,
  \dodoi{10.1051/0004-6361/201423723}

\bibitem[{Currie {et~al.}(2023)Currie, Meadows, \& Rasmussen}]{currie2023there}
Currie, M.~H., Meadows, V.~S., \& Rasmussen, K.~C. 2023, The Planetary Science
  Journal, 4, 83

\bibitem[{da~Costa-Luis(2019)}]{da2019tqdm}
da~Costa-Luis, C. 2019, Journal of Open Source Software, 4, 1277

\bibitem[{de~Kok {et~al.}(2013)de~Kok, Brogi, Snellen, Birkby, Albrecht, \&
  de~Mooij}]{de2013detection}
de~Kok, R.~J., Brogi, M., Snellen, I.~A., {et~al.} 2013, Astronomy \&
  Astrophysics, 554, A82

\bibitem[{Ehrenreich {et~al.}(2020)Ehrenreich, Lovis, Allart, Osorio, Pepe,
  Cristiani, Rebolo, Santos, Borsa, Demangeon,
  {et~al.}}]{ehrenreich2020nightside}
Ehrenreich, D., Lovis, C., Allart, R., {et~al.} 2020, Nature, 580, 597

\bibitem[{Finnerty {et~al.}(2021)Finnerty, Buzard, Pelletier, Piskorz,
  Lockwood, Bender, Benneke, \& Blake}]{finnerty2021contrast}
Finnerty, L., Buzard, C., Pelletier, S., {et~al.} 2021, The Astronomical
  Journal, 161, 104

\bibitem[{Flowers {et~al.}(2019)Flowers, Brogi, Rauscher, Kempton, \&
  Chiavassa}]{flowers2019high}
Flowers, E., Brogi, M., Rauscher, E., Kempton, E. M.-R., \& Chiavassa, A. 2019,
  The Astronomical Journal, 157, 209

\bibitem[{Foreman-Mackey(2016)}]{corner}
Foreman-Mackey, D. 2016, The Journal of Open Source Software, 1, 24,
  \dodoi{10.21105/joss.00024}

\bibitem[{Foreman-Mackey {et~al.}(2021)Foreman-Mackey, Luger, Agol, Barclay,
  Bouma, Brandt, Czekala, David, Dong, Gilbert, Gordon, Hedges, Hey, Morris,
  Price-Whelan, \& Savel}]{foreman_mackey_daniel_2021_7191939}
Foreman-Mackey, D., Luger, R., Agol, E., {et~al.} 2021, {exoplanet:
  Gradient-based probabilistic inference for exoplanet data \& other
  astronomical time series}, 0.5.1,  Zenodo, \dodoi{10.5281/zenodo.7191939}

\bibitem[{Gabri{\'e} {et~al.}(2022)Gabri{\'e}, Rotskoff, \&
  Vanden-Eijnden}]{gabrie2022adaptive}
Gabri{\'e}, M., Rotskoff, G.~M., \& Vanden-Eijnden, E. 2022, Proceedings of the
  National Academy of Sciences, 119, e2109420119

\bibitem[{Gandhi {et~al.}(2022)Gandhi, Kesseli, Snellen, Brogi, Wardenier,
  Parmentier, Welbanks, \& Savel}]{gandhi2022spatially}
Gandhi, S., Kesseli, A., Snellen, I., {et~al.} 2022, Monthly Notices of the
  Royal Astronomical Society

\bibitem[{Gandhi {et~al.}(2019)Gandhi, Madhusudhan, Hawker, \&
  Piette}]{gandhi2019hydra}
Gandhi, S., Madhusudhan, N., Hawker, G., \& Piette, A. 2019, The Astronomical
  Journal, 158, 228

\bibitem[{Gandhi {et~al.}(2023)Gandhi, Kesseli, Zhang, Louca, Snellen, Brogi,
  Miguel, Casasayas-Barris, Pelletier, Landman, {et~al.}}]{gandhi2023retrieval}
Gandhi, S., Kesseli, A., Zhang, Y., {et~al.} 2023, The Astronomical Journal,
  165, 242

\bibitem[{Gao {et~al.}(2021)Gao, Wakeford, Moran, \&
  Parmentier}]{gao2021aerosols}
Gao, P., Wakeford, H.~R., Moran, S.~E., \& Parmentier, V. 2021, Aerosols in
  exoplanet atmospheres,  Wiley Online Library

\bibitem[{Gardner {et~al.}(2006)Gardner, Mather, Clampin, Doyon, Greenhouse,
  Hammel, Hutchings, Jakobsen, Lilly, Long, {et~al.}}]{gardner2006james}
Gardner, J.~P., Mather, J.~C., Clampin, M., {et~al.} 2006, Space Science
  Reviews, 123, 485

\bibitem[{Gelman {et~al.}(2004)Gelman, Carlin, Stern, \&
  Rubin}]{gelman2004bayesian}
Gelman, A., Carlin, J.~B., Stern, H.~S., \& Rubin, D.~B. 2004, CRC Texts in
  Statistical Science, 136

\bibitem[{Gelman \& Rubin(1992)}]{gelman1992inference}
Gelman, A., \& Rubin, D.~B. 1992, Statistical science, 7, 457

\bibitem[{Giacobbe {et~al.}(2021)Giacobbe, Brogi, Gandhi, Cubillos, Bonomo,
  Sozzetti, Fossati, Guilluy, Carleo, Rainer, {et~al.}}]{giacobbe2021five}
Giacobbe, P., Brogi, M., Gandhi, S., {et~al.} 2021, Nature, 592, 205

\bibitem[{Gibson {et~al.}(2022)Gibson, Nugroho, Lothringer, Maguire, \&
  Sing}]{gibson2022relative}
Gibson, N.~P., Nugroho, S.~K., Lothringer, J., Maguire, C., \& Sing, D.~K.
  2022, Monthly Notices of the Royal Astronomical Society, 512, 4618

\bibitem[{Gibson {et~al.}(2020)Gibson, Merritt, Nugroho, Cubillos, de~Mooij,
  Mikal-Evans, Fossati, Lothringer, Nikolov, Sing,
  {et~al.}}]{gibson2020detection}
Gibson, N.~P., Merritt, S., Nugroho, S.~K., {et~al.} 2020, Monthly Notices of
  the Royal Astronomical Society, 493, 2215

\bibitem[{Gray(2021)}]{gray2021observation}
Gray, D.~F. 2021, The observation and analysis of stellar photospheres
  (Cambridge university press)

\bibitem[{Greene {et~al.}(2016)Greene, Line, Montero, Fortney, Lustig-Yaeger,
  \& Luther}]{greene2016characterizing}
Greene, T.~P., Line, M.~R., Montero, C., {et~al.} 2016, The Astrophysical
  Journal, 817, 17

\bibitem[{Grenander \& Miller(1994)}]{grenander1994representations}
Grenander, U., \& Miller, M.~I. 1994, Journal of the Royal Statistical Society:
  Series B (Methodological), 56, 549

\bibitem[{Gully-Santiago \& Morley(2022)}]{gully2022interpretable}
Gully-Santiago, M., \& Morley, C.~V. 2022, The Astrophysical Journal, 941, 200

\bibitem[{Harris {et~al.}(2020)Harris, Millman, van~der Walt, Gommers,
  Virtanen, Cournapeau, Wieser, Taylor, Berg, Smith, Kern, Picus, Hoyer, van
  Kerkwijk, Brett, Haldane, Fernández~del Río, Wiebe, Peterson,
  Gérard-Marchant, Sheppard, Reddy, Weckesser, Abbasi, Gohlke, \&
  Oliphant}]{2020NumPy-Array}
Harris, C.~R., Millman, K.~J., van~der Walt, S.~J., {et~al.} 2020, Nature, 585,
  357–362, \dodoi{10.1038/s41586-020-2649-2}

\bibitem[{Hawker {et~al.}(2018)Hawker, Madhusudhan, Cabot, \&
  Gandhi}]{hawker2018evidence}
Hawker, G.~A., Madhusudhan, N., Cabot, S.~H., \& Gandhi, S. 2018, The
  Astrophysical Journal Letters, 863, L11

\bibitem[{Hunter(2007)}]{hunter2007matplotlib}
Hunter, J.~D. 2007, Computing in science \& engineering, 9, 90

\bibitem[{Husser {et~al.}(2013)Husser, Wende-von Berg, Dreizler, Homeier,
  Reiners, Barman, \& Hauschildt}]{husser2013new}
Husser, T.-O., Wende-von Berg, S., Dreizler, S., {et~al.} 2013, Astronomy \&
  Astrophysics, 553, A6

\bibitem[{Jolliffe \& Cadima(2016)}]{jolliffe2016principal}
Jolliffe, I.~T., \& Cadima, J. 2016, Philosophical Transactions of the Royal
  Society A: Mathematical, Physical and Engineering Sciences, 374, 20150202

\bibitem[{Kanumalla {et~al.}(2024)Kanumalla, Line, Mansfield, Welbanks, Smith,
  Bean, Pino, Brogi, \& Panwar}]{kanumalla2024igrins}
Kanumalla, K., Line, M.~R., Mansfield, M.~W., {et~al.} 2024, arXiv preprint
  arXiv:2406.14072

\bibitem[{Kasper {et~al.}(2021)Kasper, Bean, Line, Seifahrt, St{\"u}rmer, Pino,
  D{\'e}sert, \& Brogi}]{kasper2021confirmation}
Kasper, D., Bean, J.~L., Line, M.~R., {et~al.} 2021, The Astrophysical Journal
  Letters, 921, L18

\bibitem[{Kausch {et~al.}(2015)Kausch, Noll, Smette, Kimeswenger, Barden,
  Szyszka, Jones, Sana, Horst, \& Kerber}]{kausch2015molecfit}
Kausch, W., Noll, S., Smette, A., {et~al.} 2015, Astronomy \& Astrophysics,
  576, A78

\bibitem[{Kawahara {et~al.}(2022)Kawahara, Kawashima, Masuda, Crossfield,
  Pannier, \& van~den Bekerom}]{kawahara2022autodifferentiable}
Kawahara, H., Kawashima, Y., Masuda, K., {et~al.} 2022, The Astrophysical
  Journal Supplement Series, 258, 31

\bibitem[{Kempton \& Rauscher(2012)}]{kempton2012constraining}
Kempton, E. M.-R., \& Rauscher, E. 2012, The Astrophysical Journal, 751, 117

\bibitem[{Kesseli {et~al.}(2022)Kesseli, Snellen, Casasayas-Barris,
  Molli{\`e}re, \& S{\'a}nchez-L{\'o}pez}]{kesseli2022atomic}
Kesseli, A.~Y., Snellen, I., Casasayas-Barris, N., Molli{\`e}re, P., \&
  S{\'a}nchez-L{\'o}pez, A. 2022, The Astronomical Journal, 163, 107

\bibitem[{Kobyzev {et~al.}(2020)Kobyzev, Prince, \&
  Brubaker}]{kobyzev2020normalizing}
Kobyzev, I., Prince, S.~J., \& Brubaker, M.~A. 2020, IEEE transactions on
  pattern analysis and machine intelligence, 43, 3964

\bibitem[{Lam {et~al.}(2015)Lam, Pitrou, \& Seibert}]{lam2015numba}
Lam, S.~K., Pitrou, A., \& Seibert, S. 2015, in Proceedings of the Second
  Workshop on the LLVM Compiler Infrastructure in HPC, 1--6

\bibitem[{Langeveld {et~al.}(2021)Langeveld, Madhusudhan, Cabot, \&
  Hodgkin}]{langeveld2021assessing}
Langeveld, A.~B., Madhusudhan, N., Cabot, S.~H., \& Hodgkin, S.~T. 2021,
  Monthly Notices of the Royal Astronomical Society, 502, 4392

\bibitem[{Le {et~al.}(2015)Le, Pak, Jaffe, Kaplan, Lee, Im, \&
  Seifahrt}]{le2015exposure}
Le, H. A.~N., Pak, S., Jaffe, D.~T., {et~al.} 2015, Advances in Space Research,
  55, 2509

\bibitem[{Lee {et~al.}(2022)Lee, Wardenier, Prinoth, Parmentier, Grimm,
  Baeyens, Carone, Christie, Deitrick, Kitzmann, {et~al.}}]{lee20223d}
Lee, E.~K., Wardenier, J.~P., Prinoth, B., {et~al.} 2022, The Astrophysical
  Journal, 929, 180

\bibitem[{{Lee} {et~al.}(2017){Lee}, {Gullikson}, \& {Kaplan}}]{igrins2017ppl}
{Lee}, J.-J., {Gullikson}, K., \& {Kaplan}, K. 2017, {Igrins/Plp 2.2.0},
  Zenodo,  Zenodo, \dodoi{10.5281/zenodo.845059}

\bibitem[{Line {et~al.}(2014)Line, Knutson, Wolf, \& Yung}]{line2014systematic}
Line, M.~R., Knutson, H., Wolf, A.~S., \& Yung, Y.~L. 2014, The Astrophysical
  Journal, 783, 70

\bibitem[{Line {et~al.}(2015)Line, Teske, Burningham, Fortney, \&
  Marley}]{line2015uniform}
Line, M.~R., Teske, J., Burningham, B., Fortney, J.~J., \& Marley, M.~S. 2015,
  The Astrophysical Journal, 807, 183

\bibitem[{Line \& Yung(2013)}]{line2013systematiciii}
Line, M.~R., \& Yung, Y.~L. 2013, The Astrophysical Journal, 779, 3

\bibitem[{Line {et~al.}(2013)Line, Wolf, Zhang, Knutson, Kammer, Ellison,
  Deroo, Crisp, \& Yung}]{line2013systematic}
Line, M.~R., Wolf, A.~S., Zhang, X., {et~al.} 2013, The Astrophysical Journal,
  775, 137

\bibitem[{Line {et~al.}(2016)Line, Stevenson, Bean, Desert, Fortney, Kreidberg,
  Madhusudhan, Showman, \& Diamond-Lowe}]{line2016no}
Line, M.~R., Stevenson, K.~B., Bean, J., {et~al.} 2016, The Astronomical
  Journal, 152, 203

\bibitem[{Line {et~al.}(2017)Line, Marley, Liu, Burningham, Morley, Hinkel,
  Teske, Fortney, Freedman, \& Lupu}]{line2017uniform}
Line, M.~R., Marley, M.~S., Liu, M.~C., {et~al.} 2017, The Astrophysical
  Journal, 848, 83

\bibitem[{Line {et~al.}(2021)Line, Brogi, Bean, Gandhi, Zalesky, Parmentier,
  Smith, Mace, Mansfield, Kempton, {et~al.}}]{line2021solar}
Line, M.~R., Brogi, M., Bean, J.~L., {et~al.} 2021, Nature, 598, 580

\bibitem[{Lockwood {et~al.}(2014)Lockwood, Johnson, Bender, Carr, Barman,
  Richert, \& Blake}]{lockwood2014near}
Lockwood, A.~C., Johnson, J.~A., Bender, C.~F., {et~al.} 2014, The
  Astrophysical Journal Letters, 783, L29

\bibitem[{Lopez-Morales {et~al.}(2019)Lopez-Morales, Ben-Ami, Gonzalez-Abad,
  Garcia-Mejia, Dietrich, \& Szentgyorgyi}]{lopez2019optimizing}
Lopez-Morales, M., Ben-Ami, S., Gonzalez-Abad, G., {et~al.} 2019, The
  Astronomical Journal, 158, 24

\bibitem[{Lord(1992)}]{lord1992new}
Lord, S.~D. 1992, A new software tool for computing Earth's atmospheric
  transmission of near-and far-infrared radiation, Vol. 103957 (Ames Research
  Center)

\bibitem[{{Mace} {et~al.}(2018){Mace}, {Sokal}, {Lee}, {Oh}, {Park}, {Lee},
  {Good}, {MacQueen}, {Oh}, {Kaplan}, {Kidder}, {Chun}, {Yuk}, {Jeong}, {Pak},
  {Kim}, {Nah}, {Lee}, {Yu}, {Hwang}, {Park}, {Kim}, {Chinn}, {Peck}, {Diaz},
  {Rutten}, {Prato}, {Jacoby}, {Cornelius}, {Hardesty}, {DeGroff}, {Dunham},
  {Levine}, {Nofi}, {Lopez-Valdivia}, {Weinberger}, \&
  {Jaffe}}]{mace2018igrins}
{Mace}, G., {Sokal}, K., {Lee}, J.-J., {et~al.} 2018, in Society of
  Photo-Optical Instrumentation Engineers (SPIE) Conference Series, Vol. 10702,
  Ground-based and Airborne Instrumentation for Astronomy VII, ed. C.~J.
  {Evans}, L.~{Simard}, \& H.~{Takami}, 107020Q, \dodoi{10.1117/12.2312345}

\bibitem[{Madhusudhan(2018)}]{madhusudhan2018atmospheric}
Madhusudhan, N. 2018, arXiv preprint arXiv:1808.04824

\bibitem[{Madhusudhan(2019)}]{madhusudhan2019exoplanetary}
---. 2019, arXiv preprint arXiv:1904.03190

\bibitem[{Madhusudhan \& Seager(2009)}]{madhusudhan2009temperature}
Madhusudhan, N., \& Seager, S. 2009, The Astrophysical Journal, 707, 24

\bibitem[{Maguire {et~al.}(2022)Maguire, Gibson, Nugroho, Ramkumar, Fortune,
  Merritt, \& de~Mooij}]{maguire2022high}
Maguire, C., Gibson, N.~P., Nugroho, S.~K., {et~al.} 2022, Monthly Notices of
  the Royal Astronomical Society

\bibitem[{Mangoubi {et~al.}(2018)Mangoubi, Pillai, \& Smith}]{mangoubi2018does}
Mangoubi, O., Pillai, N.~S., \& Smith, A. 2018, arXiv preprint arXiv:1808.03230

\bibitem[{Margossian(2019)}]{margossian2019review}
Margossian, C.~C. 2019, Wiley interdisciplinary reviews: data mining and
  knowledge discovery, 9, e1305

\bibitem[{Maxted {et~al.}(2012)Maxted, Anderson, Cameron, Doyle, Fumel, Gillon,
  Hellier, Jehin, Lendl, Pepe, {et~al.}}]{maxted2012wasp}
Maxted, P., Anderson, D., Cameron, A.~C., {et~al.} 2012, Publications of the
  Astronomical Society of the Pacific, 125, 48

\bibitem[{Mbarek \& Kempton(2016)}]{mbarek2016clouds}
Mbarek, R., \& Kempton, E. M.-R. 2016, The Astrophysical Journal, 827, 121

\bibitem[{{M}c{K}inney(2010)}]{mckinney2010data}
{M}c{K}inney, W. 2010, in {P}roceedings of the 9th {P}ython in {S}cience
  {C}onference, ed. {S}t\'efan van~der {W}alt \& {J}arrod {M}illman, 56 -- 61,
  \dodoi{10.25080/Majora-92bf1922-00a}

\bibitem[{Meech {et~al.}(2022)Meech, Aigrain, Brogi, \&
  Birkby}]{meech2022applications}
Meech, A., Aigrain, S., Brogi, M., \& Birkby, J.~L. 2022, Monthly Notices of
  the Royal Astronomical Society, 512, 2604

\bibitem[{Neal(2003)}]{neal2003slice}
Neal, R.~M. 2003, The annals of statistics, 31, 705

\bibitem[{Nortmann {et~al.}(2024)Nortmann, Lesjak, Yan, Cont, Czesla, Lavail,
  Rains, Nagel, Boldt-Christmas, Hatzes, {et~al.}}]{nortmann2024crires}
Nortmann, L., Lesjak, F., Yan, F., {et~al.} 2024, arXiv preprint
  arXiv:2404.12363

\bibitem[{Papamakarios {et~al.}(2021)Papamakarios, Nalisnick, Rezende, Mohamed,
  \& Lakshminarayanan}]{papamakarios2021normalizing}
Papamakarios, G., Nalisnick, E., Rezende, D.~J., Mohamed, S., \&
  Lakshminarayanan, B. 2021, The Journal of Machine Learning Research, 22, 2617

\bibitem[{{Park} {et~al.}(2014){Park}, {Jaffe}, {Yuk}, {Chun}, {Pak}, {Kim},
  {Pavel}, {Lee}, {Oh}, {Jeong}, {Sim}, {Lee}, {Nguyen Le}, {Strubhar},
  {Gully-Santiago}, {Oh}, {Cha}, {Moon}, {Park}, {Brooks}, {Ko}, {Han}, {Nah},
  {Hill}, {Lee}, {Barnes}, {Yu}, {Kaplan}, {Mace}, {Kim}, {Lee}, {Hwang}, \&
  {Park}}]{park2014igrins}
{Park}, C., {Jaffe}, D.~T., {Yuk}, I.-S., {et~al.} 2014, in Society of
  Photo-Optical Instrumentation Engineers (SPIE) Conference Series, Vol. 9147,
  Ground-based and Airborne Instrumentation for Astronomy V, ed. S.~K.
  {Ramsay}, I.~S. {McLean}, \& H.~{Takami}, 91471D, \dodoi{10.1117/12.2056431}

\bibitem[{P{\'e}rez \& Granger(2007)}]{perez2007ipython}
P{\'e}rez, F., \& Granger, B.~E. 2007, Computing in Science \& Engineering, 9,
  21

\bibitem[{Pino {et~al.}(2020)Pino, D{\'e}sert, Brogi, Malavolta, Wyttenbach,
  Line, Hoeijmakers, Fossati, Bonomo, Nascimbeni, {et~al.}}]{pino2020neutral}
Pino, L., D{\'e}sert, J.-M., Brogi, M., {et~al.} 2020, The Astrophysical
  Journal Letters, 894, L27

\bibitem[{Polyansky {et~al.}(2018)Polyansky, Kyuberis, Zobov, Tennyson,
  Yurchenko, \& Lodi}]{polyansky2018exomol}
Polyansky, O.~L., Kyuberis, A.~A., Zobov, N.~F., {et~al.} 2018, Monthly Notices
  of the Royal Astronomical Society, 480, 2597

\bibitem[{Powell {et~al.}(2018)Powell, Zhang, Gao, \&
  Parmentier}]{powell2018formation}
Powell, D., Zhang, X., Gao, P., \& Parmentier, V. 2018, The Astrophysical
  Journal, 860, 18

\bibitem[{{Price-Whelan} {et~al.}(2018){Price-Whelan}, {Sip{\H{o}}cz},
  {G{\"u}nther}, {Lim}, {Crawford}, {Conseil}, {Shupe}, {Craig}, {Dencheva},
  {Ginsburg}, {VanderPlas}, {Bradley}, {P{\'e}rez-Su{\'a}rez}, {de Val-Borro},
  {Paper Contributors}, {Aldcroft}, {Cruz}, {Robitaille}, {Tollerud},
  {Coordination Committee}, {Ardelean}, {Babej}, {Bach}, {Bachetti}, {Bakanov},
  {Bamford}, {Barentsen}, {Barmby}, {Baumbach}, {Berry}, {Biscani}, {Boquien},
  {Bostroem}, {Bouma}, {Brammer}, {Bray}, {Breytenbach}, {Buddelmeijer},
  {Burke}, {Calderone}, {Cano Rodr{\'\i}guez}, {Cara}, {Cardoso}, {Cheedella},
  {Copin}, {Corrales}, {Crichton}, {D{\textquoteright}Avella}, {Deil},
  {Depagne}, {Dietrich}, {Donath}, {Droettboom}, {Earl}, {Erben}, {Fabbro},
  {Ferreira}, {Finethy}, {Fox}, {Garrison}, {Gibbons}, {Goldstein}, {Gommers},
  {Greco}, {Greenfield}, {Groener}, {Grollier}, {Hagen}, {Hirst}, {Homeier},
  {Horton}, {Hosseinzadeh}, {Hu}, {Hunkeler}, {Ivezi{\'c}}, {Jain}, {Jenness},
  {Kanarek}, {Kendrew}, {Kern}, {Kerzendorf}, {Khvalko}, {King}, {Kirkby},
  {Kulkarni}, {Kumar}, {Lee}, {Lenz}, {Littlefair}, {Ma}, {Macleod},
  {Mastropietro}, {McCully}, {Montagnac}, {Morris}, {Mueller}, {Mumford},
  {Muna}, {Murphy}, {Nelson}, {Nguyen}, {Ninan}, {N{\"o}the}, {Ogaz}, {Oh},
  {Parejko}, {Parley}, {Pascual}, {Patil}, {Patil}, {Plunkett}, {Prochaska},
  {Rastogi}, {Reddy Janga}, {Sabater}, {Sakurikar}, {Seifert}, {Sherbert},
  {Sherwood-Taylor}, {Shih}, {Sick}, {Silbiger}, {Singanamalla}, {Singer},
  {Sladen}, {Sooley}, {Sornarajah}, {Streicher}, {Teuben}, {Thomas},
  {Tremblay}, {Turner}, {Terr{\'o}n}, {van Kerkwijk}, {de la Vega}, {Watkins},
  {Weaver}, {Whitmore}, {Woillez}, {Zabalza}, \& {Contributors}}]{astropy:2018}
{Price-Whelan}, A.~M., {Sip{\H{o}}cz}, B.~M., {G{\"u}nther}, H.~M., {et~al.}
  2018, \aj, 156, 123, \dodoi{10.3847/1538-3881/aabc4f}

\bibitem[{Prinoth {et~al.}(2022)Prinoth, Hoeijmakers, Kitzmann, Sandvik,
  Seidel, Lendl, Borsato, Thorsbro, Anderson, Barrado,
  {et~al.}}]{prinoth2022titanium}
Prinoth, B., Hoeijmakers, H.~J., Kitzmann, D., {et~al.} 2022, Nature Astronomy,
  6, 449

\bibitem[{Savel {et~al.}(2021)Savel, Kempton, Malik, Komacek, Bean, May,
  Stevenson, Mansfield, \& Rauscher}]{savel2021no}
Savel, A.~B., Kempton, E. M.-R., Malik, M., {et~al.} 2021, arXiv preprint
  arXiv:2109.00163

\bibitem[{Showman \& Guillot(2002)}]{showman2002atmospheric}
Showman, A., \& Guillot, T. 2002, Astronomy \& Astrophysics, 385, 166

\bibitem[{Showman {et~al.}(2013)Showman, Fortney, Lewis, \&
  Shabram}]{showman2013doppler}
Showman, A.~P., Fortney, J.~J., Lewis, N.~K., \& Shabram, M. 2013, The
  Astrophysical Journal, 762, 24

\bibitem[{Smette {et~al.}(2015)Smette, Sana, Noll, Horst, Kausch, Kimeswenger,
  Barden, Szyszka, Jones, Gallenne, {et~al.}}]{smette2015molecfit}
Smette, A., Sana, H., Noll, S., {et~al.} 2015, Astronomy \& Astrophysics, 576,
  A77

\bibitem[{Smith {et~al.}(2024)Smith, Line, Bean, Brogi, August, Welbanks,
  Desert, Lunine, Sanchez, Mansfield, {et~al.}}]{smith2024combined}
Smith, P.~C., Line, M.~R., Bean, J.~L., {et~al.} 2024, The Astronomical
  Journal, 167, 110

\bibitem[{Snellen {et~al.}(2010)Snellen, De~Kok, De~Mooij, \&
  Albrecht}]{snellen2010orbital}
Snellen, I.~A., De~Kok, R.~J., De~Mooij, E.~J., \& Albrecht, S. 2010, Nature,
  465, 1049

\bibitem[{Tan \& Komacek(2019)}]{tan2019atmospheric}
Tan, X., \& Komacek, T.~D. 2019, The Astrophysical Journal, 886, 26

\bibitem[{Tierney(1994)}]{tierney1994markov}
Tierney, L. 1994, the Annals of Statistics, 1701

\bibitem[{Ulrich~K{\"a}ufl(2008)}]{ulrich2008crires}
Ulrich~K{\"a}ufl, H. 2008, in Precision Spectroscopy in Astrophysics:
  Proceedings of the ESO/Lisbon/Aveiro Conference held in Aveiro, Portugal,
  11--15 September 2006, Springer, 227--230

\bibitem[{van Sluijs {et~al.}(2022)van Sluijs, Birkby, Lothringer, Lee,
  Crossfield, Parmentier, Brogi, Kulesa, McCarthy, Powell,
  {et~al.}}]{van2022carbon}
van Sluijs, L., Birkby, J.~L., Lothringer, J., {et~al.} 2022, arXiv preprint
  arXiv:2203.13234

\bibitem[{Virtanen {et~al.}(2020)Virtanen, Gommers, Oliphant, Haberland, Reddy,
  Cournapeau, Burovski, Peterson, Weckesser, Bright,
  {et~al.}}]{virtanen2020scipy}
Virtanen, P., Gommers, R., Oliphant, T.~E., {et~al.} 2020, Nature methods, 17,
  261

\bibitem[{Wang {et~al.}(2022)Wang, Latouf, Plavchan, Cale, Blake, Artigau,
  Lisse, Gagn{\'e}, Crass, \& Tanner}]{wang2022characterizing}
Wang, S.~X., Latouf, N., Plavchan, P., {et~al.} 2022, The Astronomical Journal,
  164, 211

\bibitem[{Wardenier {et~al.}(2021)Wardenier, Parmentier, Lee, Line, \&
  Gharib-Nezhad}]{wardenier2021decomposing}
Wardenier, J.~P., Parmentier, V., Lee, E.~K., Line, M., \& Gharib-Nezhad, E.
  2021, arXiv preprint arXiv:2105.11034

\bibitem[{Wardenier {et~al.}(2023)Wardenier, Parmentier, Line, \&
  Lee}]{wardenier2023modelling}
Wardenier, J.~P., Parmentier, V., Line, M.~R., \& Lee, E.~K. 2023, arXiv
  preprint arXiv:2307.04931

\bibitem[{Wong {et~al.}(2023{\natexlab{a}})Wong, Isi, \&
  Edwards}]{wong2023fast}
Wong, K.~W., Isi, M., \& Edwards, T.~D. 2023{\natexlab{a}}, arXiv preprint
  arXiv:2302.05333

\bibitem[{{Wong} {et~al.}(2022){Wong}, {Gabri{\'e}}, \&
  {Foreman-Mackey}}]{wong2022flow}
{Wong}, K. W.~K., {Gabri{\'e}}, M., \& {Foreman-Mackey}, D. 2022, arXiv
  e-prints, arXiv:2211.06397.
\newblock \doarXiv{2211.06397}

\bibitem[{Wong {et~al.}(2023{\natexlab{b}})Wong, Gabri\'e, \&
  Foreman-Mackey}]{Wong:2022xvh}
Wong, K. W.~k., Gabri\'e, M., \& Foreman-Mackey, D. 2023{\natexlab{b}}, J. Open
  Source Softw., 8, 5021, \dodoi{10.21105/joss.05021}

\bibitem[{Xue {et~al.}(2024)Xue, Bean, Zhang, Welbanks, Lunine, \&
  August}]{xue2024jwst}
Xue, Q., Bean, J.~L., Zhang, M., {et~al.} 2024, The Astrophysical Journal
  Letters, 963, L5

\bibitem[{{Yuk} {et~al.}(2010){Yuk}, {Jaffe}, {Barnes}, {Chun}, {Park}, {Lee},
  {Lee}, {Wang}, {Park}, {Pak}, {Strubhar}, {Deen}, {Oh}, {Seo}, {Pyo}, {Park},
  {Lacy}, {Goertz}, {Rand}, \& {Gully-Santiago}}]{yuk2010igrins}
{Yuk}, I.-S., {Jaffe}, D.~T., {Barnes}, S., {et~al.} 2010, in Society of
  Photo-Optical Instrumentation Engineers (SPIE) Conference Series, Vol. 7735,
  Ground-based and Airborne Instrumentation for Astronomy III, ed. I.~S.
  {McLean}, S.~K. {Ramsay}, \& H.~{Takami}, 77351M, \dodoi{10.1117/12.856864}

\bibitem[{Zhang {et~al.}(2017)Zhang, Kempton, \&
  Rauscher}]{zhang2017constraining}
Zhang, J., Kempton, E. M.-R., \& Rauscher, E. 2017, The Astrophysical Journal,
  851, 84

\bibitem[{Zhang {et~al.}(2020)Zhang, Chachan, Kempton, Knutson,
  {et~al.}}]{zhang2020platon}
Zhang, M., Chachan, Y., Kempton, E. M.-R., Knutson, H.~A., {et~al.} 2020, The
  Astrophysical Journal, 899, 27

\end{thebibliography}
\bibliographystyle{aasjournal}

\end{document}